\documentclass[aps,prb,reprint,twocolumn,superscriptaddress,floatfix,nofootinbib]{revtex4-1}
\usepackage{amssymb}
\usepackage{amsmath,mathtools}
\usepackage{braket}
\usepackage[version=3]{mhchem}
\usepackage{amssymb}
\usepackage{amsmath,mathtools}
\usepackage{braket}
\usepackage[version=3]{mhchem}
\usepackage{color}
\usepackage{tikz}
\usepackage{tikzscale}
\usepackage{graphicx}
\graphicspath{{./figures/}}
\usepackage{dcolumn}
\usepackage{bm}
\tikzstyle{tensor}=[rectangle,draw=blue!50,fill=blue!20,thick]
\definecolor{jadclr}{rgb}{0,0.5,0}
\definecolor{jadcolor}{rgb}{0.5,0,0}
\usepackage[bookmarks=true,colorlinks,linkcolor=jadclr,urlcolor=jadcolor,citecolor=jadcolor]{hyperref}

\newcommand{\norm}[1]{\lVert#1\rVert}

\newcommand{\pll}[2]{\frac{\partial #1}{\partial #2}}

\begin{document}

\title{
    Hybrid infinite time-evolving block decimation algorithm for long-range multi-dimensional quantum many-body systems
} 



\author{Tomohiro Hashizume}
\affiliation{Department of Physics and SUPA, University of Strathclyde, Glasgow G4 0NG, United Kingdom}

\author{Jad C.~Halimeh}
\affiliation{Kirchhoff Institute for Physics, Heidelberg University, 69120 Heidelberg, Germany}
\affiliation{Institute for Theoretical Physics, Heidelberg University, 69120 Heidelberg, Germany}
\affiliation{Max Planck Institute for the Physics of Complex Systems, 01187 Dresden, Germany}

\author{Ian P.~McCulloch}
\affiliation{School of Mathematics and Physics, The University of Queensland, St.~Lucia, QLD 4072, Australia}

\date{\today}

\begin{abstract}
In recent years, the infinite time-evolution block decimation (iTEBD) method has been demonstrated to be one of the most efficient and powerful numerical schemes for time-evolution in one-dimensional quantum many-body systems. However, a major shortcoming of the method, along with other state-of-the-art algorithms for many-body dynamics, has been their restriction to one spatial dimension. We present an algorithm based on a \textit{hybrid} extension of iTEBD where finite blocks of a chain are first locally time-evolved before an iTEBD-like method combines these processes globally. This in turn permits simulating the dynamics of many-body systems in the thermodynamic limit in $d\geq1$ dimensions including in the presence of long-range interactions. Our work paves the way for simulating the dynamics of many-body phenomena that occur exclusively in higher dimensions, and whose numerical treatments have hitherto been limited to exact diagonalization of small systems, which fundamentally limits a proper investigation of dynamical criticality. We expect the algorithm presented here to be of significant importance to validating and guiding investigations in state-of-the-art ion-trap and ultracold-atom experiments.
\end{abstract}


\maketitle 

\section{Introduction}
\label{Introduction}
Over the past century, a lot of research in condensed matter physics has focused on exploring, understanding, and engineering exotic phenomena emerging from the interactions of many particles whose individual behavior falls short of the richness contained in that of their collective gestalt.\cite{Coleman_book} Straightforward examples of the emergent behavior of interacting microscopic particles are the distinct phases of matter due to spontaneous breaking of global symmetries, topological phase transitions and topological order, laminar flow, turbulence, and countless other phenomena the appearance of which necessarily requires the collective interactions of the microscopic constituents.\cite{Kivelson2016} 

In recent years, the study of dynamical criticality in quantum many-body systems has received huge interest.\cite{Mori2018,Zvyagin2016} In particular, two concepts of dynamical phase transitions have been at the forefront of these investigations, where both are mostly concerned with dynamics in the wake of a quench. The first relies on the dynamical behavior of a local order parameter. Criticality can then be determined by the infinite-time value of this order parameter, such that when it takes a finite value this means that the system has settled into a (global) symmetry-broken steady state, whereas a zero order parameter indicates that the system has settled into a symmetric steady state.\cite{Sciolla2010,Sciolla2011} However, criticality can also be determined by the intermediate or prethermal dynamics of the order parameter, rather than its value in the steady state. This is particularly intuitive in the case of models with no finite-temperature phase transition where a quenched system always ends up in a symmetric steady state. Nevertheless, how the order parameter evolves in time can exhibit either one of two dynamical behaviors (phases): an asymptotic decay to zero where the order parameter never changes sign, or an envelope that oscillates about zero while asymptotically decaying to it. These phases have been characterized for quenches within the ordered phase and from the ordered phase to the symmetric phase, respectively, in the nearest-neighbor quantum Ising,\cite{Calabrese2011,Calabrese2012} Bose-Hubbard,\cite{Altman2002} and XXZ\cite{Barmettler2009} chains. Recently, this notion of critical behavior in the intermediate dynamics of the order parameter has been extended to long-range interacting quantum spin chains.\cite{Halimeh2017b}

The second concept of dynamical phase transitions builds on interpreting the overlap of the time-evolved wave function of a many-body system with its initial state as a dynamical analog of the thermal partition function, with complexified evolution time standing in for inverse temperature.\cite{Heyl2013,Heyl_review} Therefore, just like critical temperatures bring about thermal phase transitions in equilibrium, critical times define so-called \textit{dynamical quantum phase transitions} (DQPT) in out-of-equilibrium many-body dynamics. The theory of DQPT has recently witnessed rapid development, with studies of its properties covering various integrable and nonintegrable systems. 
 
Nevertheless, these various phenomena require great effort to generate whether experimentally or numerically. In generic quantum many-body systems the effective Hilbert space grows exponentially with system size, rendering exact diagonalization methods simply insufficient, particularly when one is interested in critical behavior, and more so when that criticality is in the out-of-equilibrium realm. Sophisticated numerical simulations have taken on a key role in studying such systems, in particular the celebrated density matrix renormalization group (DMRG) algorithm,\cite{White1992,Schollwock2005a} and its implementation using matrix product states\cite{Uli2011} (MPS) has allowed the exploration of various equilibrium phases of matter.
Additionally, its extension to simulating the time evolution of quantum many-body systems has revealed a vast range of exotic dynamical 
properties in strongly interacting systems.\cite{Vidal2003,Daley2004,Haegeman2011,GarcIa-Ripoll2006,Zaletel2015}

At the same time, the development of high-precision experimental techniques has made possible the
realization of highly tunable quantum systems in the laboratory. Such engineered systems enable the study of not only the static properties of the underlying model, but also its dynamical properties, such as for example in the wake of quench, the parameters of which can now be very well controlled in ion-trap\cite{Jurcevic2017} and cold-atom\cite{Flaeschner2018} experiments. Moreover, state-of-the-art setups today can realize quantum many-body systems with various forms and ranges of interaction profiles and in different lattice geometries.\cite{Bloch2008a,Britton2012,Georgescu2014,Hung2016,Bentsen2019} Such \textit{quantum simulators}\cite{Martinez2016,Schweizer2019,Mil2019,Arute2019} promise to provide deep insights into the physics of both microscopic and macroscopic systems, including the study of critical dynamical phenomena in generic quantum many-body models.

To probe, verify, and calibrate a quantum simulator, the development of methods to reliably predict its properties is vital. Currently, the static properties of long-range interacting one- and two-dimensional systems in the thermodynamic limit 
are well studied using extended DMRG techniques.\cite{Verstraete2004,Saadatmand2016} In the past few years, time evolution in long-range quantum spin chains has witnessed considerable success through the time-dependent variational principle (TDVP) in uniform MPS.\cite{Haegeman2011,Haegeman2016,Zauner2018} 
The dynamics of models on finite two-dimensional lattices has also been well studied using several time-evolution 
schemes.\cite{GarcIa-Ripoll2006,Zaletel2015} However, uncovering the dynamical properties of higher-dimensional models would be the useful next step in light of the tremendous advancement in experimental setups that can now readily and reliably realize these systems. Nevertheless, numerical techniques have been restricted to just one spatial dimension when it comes to generic quantum many-body models in the thermodynamic limit. For example, numerical studies on both of the above concepts of dynamical phase transitions have been restricted to one-dimensional (1D) many-body Hamiltonians, with the exception of integrable models, such mean-field and exactly solvable free-fermionic systems, and small finite two-dimensional lattices studied in exact diagonalization where ascertaining dynamical critical behavior is necessarily inadequate so far away from the thermodynamic limit.

In this work, we introduce a new MPS algorithm that can simulate the dynamics of effectively two-dimensional (2D) systems by time-evolving mappings thereof to 1D chains with long-range fixed-length interaction profiles. This method is a combination of two techniques. The first involves local time evolution of several sites on the 1D chain using an algorithm for time-evolving a finite system. The second technique is based on iTEBD-like evolution that evolves different bulks within the 1D Hamiltonian using Suzuki-Trotter expansion. Very importantly, this method allows simulating the time evolution of long-range 1D models without assuming translational symmetry up to the size of a unit cell. By mapping 2D systems of infinite length in just a single direction to a 1D chain containing a unit cell with the appropriate interaction profile and corresponding size, our method is able to simulate the dynamics of such 2D systems and adequately determine their critical properties.

The rest of the paper is organized as follows: In Sec.~\ref{sec:mps} we provide an overview of matrix product states and operators, followed in Sec.~\ref{sec:trotter} by a description of the Suzuki-Trotter decomposition and Krylov-subspace expansion, both of which form the cornerstones of our hybrid time-evolution algorithm. In Sec.~\ref{sec:algorithm} the main features of our method~---~the hybrid infinite time-evolution block decimation (h-iTEBD) algorithm~---~are presented along with an error analysis. We explain in Sec.~\ref{sec:mapping} how one can map two-dimensional models to one-dimensional chains of fixed-length long-range interactions that are amenable for implementation using our method. Further benchmarking is presented in Sec.~\ref{sec:DQPT} in the form of a comparison of prominent DQPT results calculated using our method and other well-established techniques in 1D, and we additionally present results on DQPT in the quantum Ising model on a triangular lattice. We conclude in Sec.~\ref{sec:conc}.

\section{Matrix Product States and Matrix Product Operators}\label{sec:mps}
Matrix product states (MPS) are the representation of a state on a lattice with $L$ sites through a product of 
matrices.\cite{Schollwock2011,McCulloch2008}
Let the local state at the $i$\textsuperscript{th} site be $\sigma_{i}$, which can take any state within the local Hilbert space at site $i$. 
A component of the probability amplitude associated with a local state $\sigma_i$ is represented
by a matrix $M_i^{\sigma_{i}}$ with a bond dimension $m$.
The probability amplitude of a global state $\ket{\sigma_{1},\dots,\sigma_{L}}$ can then be represented as 
a product of all the $M_i^{\sigma_i}$ matrices.
    With this representation, a state $\ket{\Psi}$ can be represented with 
$L$ matrices with dimensions of $m \times m$ as follows:
\begin{align}\label{MPSCrude}
    \ket{\Psi} = \sum_{\mathbf{\sigma}} M^{\sigma_1}_1M^{\sigma_2}_2 \dots M^{\sigma_L}_L
    \ket{\sigma_1,\sigma_2,\dots,\sigma_L}. 
\end{align}
A product of two matrices is invariant upon substituting identity in the middle.
Therefore, one can impose a condition such as
\begin{align}
    \sum_{\sigma_l} A^{\dag \sigma_l}_l A^{\sigma_l}_l  &= 1, \\
    \sum_{\sigma_l} B^{\sigma_l}_l B^{\dag \sigma_l}_l  &= 1,
\end{align}
through some local unitary transformation.
A wavefunction can then be represented with $A$ ($B$) matrices as
\begin{align}
    \ket{\Psi} &= \sum_{\mathbf{\sigma}}  A^{\sigma_1}_1A^{\sigma_2}_2 \dots A^{\sigma_L}_L \mathbf{\Phi}
    \ket{\sigma_1,\sigma_2,\dots,\sigma_L}, \label{MPSLeftOrthogonal} \\
    \ket{\Psi} &= \sum_{\mathbf{\sigma}}  \mathbf{\Phi'} B^{\sigma_1}_1B^{\sigma_2}_2 \dots B^{\sigma_L}_L
    \ket{\sigma_1,\sigma_2,\dots,\sigma_L}, \label{MPSRightOrthogonal}
\end{align}
where $\Phi$ ($\Phi'$) is what remains after left (right) orthogonalizing all of the $M$ matrices from the left (right).

The representation in~\eqref{MPSLeftOrthogonal} can further be canonicalized using 
Vidal's $\Lambda$ $\Gamma$ notation:\cite{Vidal2007}
\begin{align}
    \ket{\Psi} = \sum_{\mathbf{\sigma}}\Lambda_1 \Gamma^{\sigma_1}_1 \Lambda_2 \Gamma^{\sigma_2}_2 \dots \Lambda_L \Gamma^{\sigma_L}_L\ket{ \sigma_1,\sigma_2,\dots,\sigma_L} ,
\end{align}
where the original $A^{\sigma_m}_{m}$ matrix corresponds to $\Lambda^{\sigma_m}_m\Gamma ^{\sigma_m}_m$.
Its infinite extension can be achieved by infinitely repeating the unit cell of length $L$.

An MPS $\ket{\Psi}$ with a bond dimension $m$ can be compressed to 
an MPS $\ket{\Psi'}$ with a lower bond dimension $m'<m$.\cite{Schollwock2011,Paeckel2019}
For $M'_i$ to be an optimal matrix that best approximates $M_i$ with lower dimension, it must satisfy 
\begin{align}
    \pll{\norm{\ket{\Psi} - \ket{\Psi'}}^2}{M'^{\sigma_i}_i} =
    \pll{\norm{\ket{\Psi} - \ket{\Psi'}}^2}{(M'^{\sigma_i}_i)^\dag} = 0 \label{ConditionForCompression}
\end{align}
for all $i$. Provided that
\begin{align}
    \begin{split}
        &\ket{\Psi}=\sum_{\sigma} \prod_{j}M_j^{\sigma_j} \ket{\sigma_1,\dots,\sigma_L}, \\
        &\ket{\Psi'}=\sum_{\sigma} \left(\prod_{j<i}A'^{\sigma_j}_j\right) M'^{\sigma_i}_i \left(\prod_{k>i}B'^{\sigma_k}_k
        \right)
        \ket{\sigma_1,\dots,\sigma_L}, \\
        &\braket{\Psi|\Psi}=\braket{\Psi'|\Psi'}=1,
    \end{split}
\end{align}
the optimum $M'^{\sigma_i}_i$ is 
\begin{align}
    M'^{\sigma_i}_i &=
    \bra{\sigma_i}
    \sum_{\sigma',\sigma_i \not\in \sigma'}
    \left(\prod_{j<i}A'^{\sigma'_j,\dag}_j\right)\bigotimes\bra{\sigma'_j}
    \left(\prod_{k>i}B'^{\sigma'_k,\dag}_k\right)\nonumber\\
    &\bigotimes\bra{\sigma'_k}
     \left( \sum_{\sigma''}
    \prod_{l} {M'_l}^{\sigma''_l} \bigotimes \ket{\sigma_l''} \right).
\end{align}

Similarly to MPS, operators on a system can be represented 
in a matrix product form.\cite{Schollwock2011,McCulloch2008,Zaletel2015,Chan2016,Pirvu2010}
This form is called a matrix product operator (MPO) and is formulated as follows.
In general, the component of an operator on a chain can be written as follows:
\begin{align}
    \hat{O} 
    = &\sum_{\sigma,\sigma'} W^{\sigma_1,\sigma'_1}W^{\sigma_2,\sigma'_2}\dots W^{\sigma_L,\sigma'_L}\nonumber\\
    &\ket{\sigma_1,\sigma_2,\dots,\sigma_L}\bra{\sigma'_1,\sigma'_2,\dots,\sigma'_L}.
    \label{MPOCrude}
\end{align}

If an operator can be expressed as a product of local operators, then most of the $W$ matrices are identity.
In such a case, MPO has a more efficient form.
When a Hamiltonian $\hat{H}$ consists of the sum of local terms, 
it can be written as a sum of parts localized to the left ($\hat{H}_{\mathrm{Left}}$)
and right ($\hat{H}_{\mathrm{Right}}$) of a boundary $\partial i$,
which resides between the $i$\textsuperscript{th} and $i+1$\textsuperscript{st} sites.
Then this Hamiltonian can be decomposed as
\begin{align}
    \hat{H} =& \hat{H}_{\mathrm{Left}} \otimes \hat{I}_{\mathrm{Right}} 
    + \hat{I}_{\mathrm{Left}} \otimes \hat{H}_{\mathrm{Right}}\nonumber\\
    &+ \sum_{l} \hat{O}_{{\mathrm{Left}},l} \otimes \hat{O}_{l} \otimes \hat{O}_{{\mathrm{Right}},l}.
    \label{HamilDecomposed}
\end{align}
Consequently, this is the equivalent of a product of the matrices of local operators.

Decomposition of $\hat{H}$ at $\partial i$ in the $i^\text{th}$ iteration looks like
\begin{align}
    \hat{H} = 
    \cdots
    \begin{pmatrix}
        \hat{I} & \cdots & \hat{O}_{i}' & \cdots & \hat{O}_{i}''\\
        & & &  \\
        0 & 0 & \hat{O}_{i}''' & \cdots & \hat{O}_{i}''''\\
        & & & \\
        0 & 0 & \cdots & 0& \hat{I}
    \end{pmatrix}_i 
    \begin{pmatrix}
        \hat{H}_{{\mathrm{Right}}} \\
        \vdots \\
        \hat{O}_{{\mathrm{Right}},i}\\
        \vdots \\
        I_{{\mathrm{Right}}}
    \end{pmatrix}_{{\mathrm{Right}}},
\end{align}
where $\hat{O}'$, $\hat{O}''$, $\hat{O}'''$, $\hat{O}''''$ are local operators that recover the summation term in~\eqref{HamilDecomposed}. This decomposition can be done until the unit cell size is reached. 
The MPO of an infinite system is then just an infinite repetition of $L$ local matrices.

\begin{figure*}[t!]
    \centering
    \includegraphics[width=0.8\textwidth]{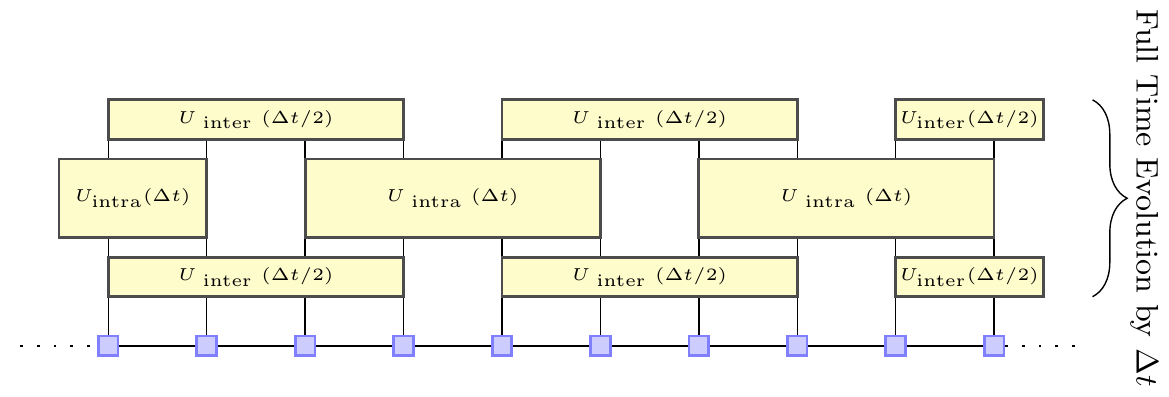}
    \caption{(Color online). Tensor network diagram of the time evolution of a state by one time step 
        using the h-iTEBD algorithm for a unit cell of four sites.
        Each square represents a matrix. The horizontal legs are the indices of a matrix and vertical 
        legs represent the local state ($\bra{\sigma}$ if going downwards, and $\ket{\sigma}$ if going upwards).
        The connected legs represent a contraction of the matrices with matrix products (horizontal) and 
        taking the overlap of the two local states (vertical).
    }
    \label{h-iTEBD_algorithm}
\end{figure*}

\section{Operator Exponentiation: Suzuki-Trotter Decomposition and Krylov Subspace Expansion}\label{sec:trotter}
An operator exponentiation of a sum of non-commuting operator is a nontrivial challenge when it comes to time-evolving systems with 
long-range interactions.
One way to perform it is by approximating the exponent of the sum to a product of exponentiated summands.
For a sum of operators with small norm, this is possible with a controlled error bound in Suzuki-Trotter decomposition.\cite{Suzuki1985,Vidal2003,Vidal2007,Orus2007,Schollwock2011}
It decomposes $\mathrm{e}^{-i\Delta t (\hat{A} + \hat{B})}$ as
\begin{align}
    \mathrm{e}^{-i\Delta t (\hat{A} + \hat{B})} = 
    \mathrm{e}^{-i\Delta t \hat{A}}\mathrm{e}^{-i\Delta t \hat{B}} + \mathcal{O}( 
    \Delta t^2).
\end{align}
This can be expanded to one higher order, without requiring any additional computational resources, as
\begin{align}
        \mathrm{e}^{-i\Delta t (\hat{A} + \hat{B})} 
        &=\mathrm{e}^{-i\frac{\Delta t}{2} \hat{A}} 
        \mathrm{e}^{-i\Delta t \hat{B}}
        \mathrm{e}^{-i\frac{\Delta t}{2} \hat{A}} + \mathcal{O}(\Delta t^3).
        \label{secondorderST}
\end{align}

This allows exponentiation to be performed individually in the subspace of the summands with an error of the order $\Delta t^3$.
Ideally, the operators $\hat{A}$ and $\hat{B}$ would be sums of commuting summands, i.e., $[\hat{A},\hat{B}]=0$.
However, this is generically not so. In the case where $[\hat{A},\hat{B}]\neq 0$, 
it is optimal that the value of the commutator be as small as possible.

After the decomposition, the exponentiated summands must be calculated.
However, exponentiation using exact diagonalization is numerically very 
expensive if not impossible for an operator with a large Hilbert space dimension.
Krylov subspace expansion approximates the operator $\hat{\mathcal{M}}$ of large Hilbert space dimension with
an operator in a small subspace spanned by the $n$ Krylov vectors
\cite{krylov1931resolution,Moler2003,GarcIa-Ripoll2006,Paeckel2019}
\begin{align}
    \left\{ \ket{\Psi(t_0)},\hat{\mathcal{M}}\ket{\Psi(t_0)}, \hat{\mathcal{M}}^2\ket{\Psi(t_0)}, \ldots,
    \hat{\mathcal{M}}^{n-1} \ket{\Psi(t_0)} \right\}\text{,}
\end{align}
where $\ket{\Psi(t_0)}$ is a Krylov vector at time $t_0$.
Once the approximated operator is found then $\exp(-\mathrm{i}\Delta t \hat{\mathcal{M}})$~|~or, in general, any function of $\hat{\mathcal{M}}$~|~can be calculated by the method of one's choice.

The Lanczos method\cite{Lanczos1950,GarcIa-Ripoll2006,Stewart2002,Orecchia2012} is one of several algorithms 
that one can use to proceed from this point. 
It tridiagonalizes an operator $\hat{\mathcal{M}}$ 
into a small operator by iteratively finding a suitable Krylov basis starting 
from $\ket{\Psi(t_0)}$.
If there is such a transformation, then $\hat{\mathcal{M}}$ can be written as follows:
\begin{align}
    \hat{\mathcal{M}} &\approx
    \begin{pmatrix}
        \alpha_0 & \beta_0 & &  \\ 
        \beta_0 & \alpha_1 & \beta_1 &  \\ 
         & \beta_1 & \alpha_2 &  \ddots &  & 0 \\
        & &\ddots & \ddots & \ddots &  & \\
        & && \ddots & \ddots &\beta_{n-3}    & \\
        & & 0 & &\beta_{n-3}   &\alpha_{n-2}  & \beta_{n-2}\\
        & &  & & &\beta_{n-2}&\alpha_{n-1}
    \end{pmatrix}.
\end{align}

Acting with $\hat{\mathcal{M}}$ on $\ket{\Psi(t_0)}$ while enforcing the orthornormalization condition $\braket{V_i|V_j}=\delta_{i,j}$
yields a recursive relation for $\alpha_i$, $\beta_i$, and the Lanczos vectors $\ket{V_i}$ as follows:
\begin{align}
    \begin{split}
        \ket{V_0} &= \ket{\Psi(t_0)}, \\
        \alpha_0 &= \bra{V_0}\hat{\mathcal{M}}\ket{V_0},\\
        \beta_0 &= \norm{\left( \hat{\mathcal{M}}-\alpha_0\right)\ket{V_0}},\\
        \alpha_i &= \bra{V_{i-1}}\hat{\mathcal{M}}\ket{V_{i-1}}, \\
        \beta_i &=\norm{\left( \hat{\mathcal{M}}-\alpha_{i}\right)\ket{V_{i}} - \beta_{i-1} \ket{V_{i-1}}}, \\
        \ket{V_{i+1}} &= \frac{1}{\beta_{i}} \left[\left( \hat{\mathcal{M}}-\alpha_{i} \right)\ket{V_{i}} - \beta_{i-1}\ket{V_{i-1}}\right].
    \end{split}
\end{align}
One can set a small parameter $\epsilon$ and stop the iteration when the value of $\beta_i$ goes below that threshold.
The tridiagonalized operator with a reduced rank can then be exponentiated using the method of one's choice.

Once the subspace is spanned by the Krylov vectors, we can write
\begin{align}
    \ket{\Psi(t_0+\Delta t)}=\exp(-i\Delta t \hat{\mathcal{M}})\ket{\Psi(t_0)} \approx \sum_{i=0}^{n-1} c_i\ket{V_i},
\end{align}
where $c_i$ is the $(i,0)$\textsuperscript{th} component of $\exp(-i\Delta t \hat{\mathcal{M}})$ in the subspace.

In general, the error from the approximation has the bound\cite{Orecchia2012,Saad,Koskela2013}
\begin{align}
    &\norm{\exp(-i\Delta t \hat{\mathcal{M}})\ket{\Psi(t_0)} - \sum_{i=0}^{n-1} c_i \ket{V_i}}  \nonumber\\
    &\leq 2\norm{\ket{\Psi(t_0)}} \frac{ \norm{(-i\Delta t\hat{\mathcal{M}})^n}}{n!}\exp(\norm{-i\Delta t \hat{\mathcal{M}}}) \text{.}
\end{align}
Therefore, the error scales as $\mathcal{O}\left( \Delta t^n \right)$. 
This error bound is only valid if $\left\{ \ket{V_i} \right\}$ are orthonormalized to infinite precision.
However, the orthonormality of $\left\{ \ket{V_i} \right\}$ 
expressed as MPS breaks down upon compression. 
This error can be reduced by introducing correction terms through adopting semiorthogonal\cite{Simon1984} or full 
reorthonormalization methods.\cite{Stewart2002}
In the case of our algorithm, such measures are unnecessary because h-iTEBD does not require Krylov subspace dimension of more than $3$. 
This is because the error from the 2\textsuperscript{nd}-order Suzuki-Trotter 
decomposition scales as $\Delta t^3$ such that the incurred accuracy from a Krylov subspace expansion larger than $3$ 
does not contribute significantly to improving the precision. 
For three Krylov vectors, the error from broken orthonormalization is effectively negligible.

\section{Algorithm}\label{sec:algorithm}
The hybrid infinite time-evolving block decimation (h-iTEBD) algorithm presented here evolves a state in time under a Hamiltonian
with translational invariance down to a length scale of $L$ sites.
Inspired by the iTEBD algorithm, h-iTEBD achieves this by breaking the Hamiltonian into two 
parts: an intra-unit cell part where the interactions are contained within the unit cell, and an inter-unit cell part where the interactions cross the unit cell boundaries.
Each part can then be evolved locally by some local time-evolution algorithm, which in our case is Krylov subspace expansion. Those separate local evolutions of inter- and intra-unit cell parts are then combined via Suzuki-Trotter iterations, as shown in Fig.~\ref{h-iTEBD_algorithm}.
Although Krylov subspace expansion is used for the local time evolution, other methods such as TDVP\cite{Haegeman2011,Haegeman2013,Haegeman2016} can equally well be used, and this is one of our future directions.
Fig.~\ref{h-iTEBD_algorithm} shows the schematic diagram of the time-evolution routine using a tensor network diagram.

We are interested in time-evolving a Hamiltonian that is translationally symmetric with unit cell size of $N$ sites. The Hamiltonian can be expressed as a sum of the component $\hat{H}_{\text{intra},i}$, 
which has only terms with interactions contained within the $i$\textsuperscript{th} unit cell, and 
a second component $\hat{H}_{\text{inter},i}$, which contains all the interactions that cross the unit cell 
boundary between the $i$\textsuperscript{th} and $i+1$\textsuperscript{st} unit cells:
\begin{equation}
    \hat{H} = \sum_{l}\big(\hat{H}_{\text{intra},l} + \hat{H}_{\text{inter},l}\big).
\end{equation}
Using Suzuki-Trotter decomposition, the time-evolution operator $ \hat{U} = \exp\left(-\mathrm{i}\Delta t \hat{H} \right)$
can be approximated as 
\begin{align}
    \hat{U} &\approx 
    \mathrm{e}^{-\mathrm{i}\frac{\Delta t}{2} \sum_{j} \hat{H}_{\text{intra},j}}
    \mathrm{e}^{-\mathrm{i}\Delta t \sum_{k} \hat{H}_{\text{inter},k}}
    \mathrm{e}^{-\mathrm{i}\frac{\Delta t}{2} \sum_{l} \hat{H}_{\text{intra},l}} \nonumber\\
    &=
    \prod_j \mathrm{e}^{-\mathrm{i}\frac{\Delta t}{2} \hat{H}_{\text{intra},j}}
    \prod_k \mathrm{e}^{-\mathrm{i}\Delta t \hat{H}_{\text{inter},k}}
    \prod_l \mathrm{e}^{-\mathrm{i}\frac{\Delta t}{2} \hat{H}_{\text{intra},l}} \nonumber\\
    &= 
    \prod_j \hat{U}_{\text{intra},j}\left(\frac{\Delta t}{2}\right)
    \prod_k \hat{U}_{\text{inter},k}\left(\Delta t \right)
    \prod_l \hat{U}_{\text{intra},l}\left(\frac{\Delta t}{2}\right).
\end{align}
$\hat{U}$ acting on a state $\ket{\Psi}$ can then be calculated with Krylov subspace expansion. 

\begin{figure}[tbh]
	\centering
	\includegraphics[width=0.95\columnwidth]{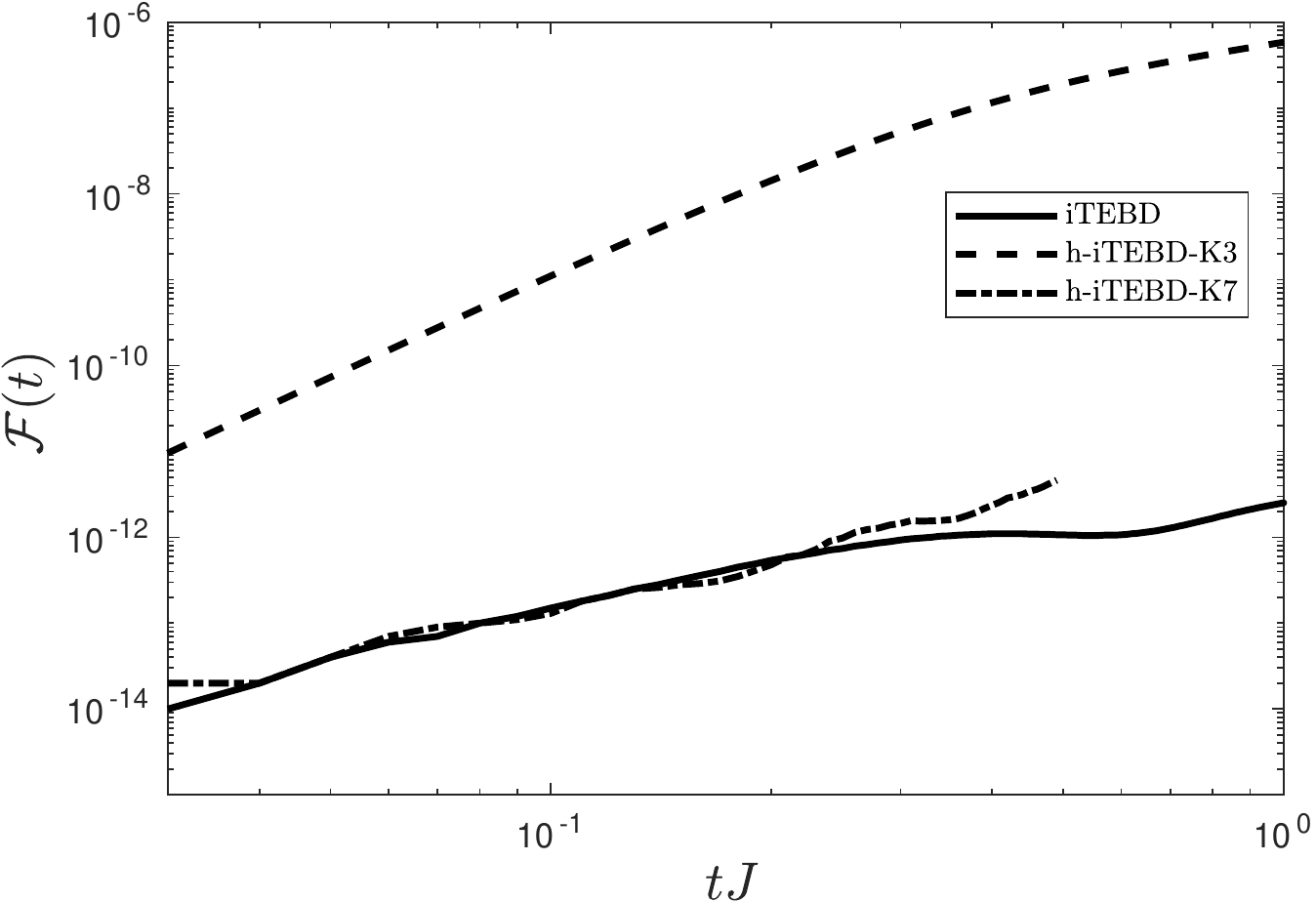}
	\caption{
		Fidelity scaling of the different local subspace dimensions for the h-iTEBD algorithm. 
		The h-iTEBD algorithm 
		with local time evolution using 3\textsuperscript{rd}-order and 7\textsuperscript{th}-order Krylov dimensions
		is compared with the quasi-exact state, which is computed using the 4\textsuperscript{th}-order iTEBD algorithm.
		The early time scaling of the Fidelity gives the expected scaling of $t^3$ and $t^2$ 
		for h-iTEBD-K3 and h-iTEBD-K7, respectively. 
		\label{Heis_overlap_error}
	}
\end{figure}

The inter- and intra-unit cell Hamiltonians have freedoms on distributing the terms that qualify to be in either one of them. 
The total error can be minimized by choosing their distribution such that 
$\left\lVert \left[ \hat{H}_{\text{inter},m},\hat{H}_{\text{intra},m} \right]\right\rVert$ and 
$\left\lVert \left[ \hat{H}_{\text{inter},m+1},\hat{H}_{\text{intra},m} \right]\right\rVert$ are smallest.
For the calculations performed in this paper, we have chosen the distribution such that the overlapping terms in  
$\hat{H}_{\text{inter},m}$ have the same contribution as the ones in $\hat{H}_{\text{intra},m}$. 
There are three contributors to the error in this method.
In the earlier time steps, these are, from largest to smallest, the error from the Suzuki-Trotter decomposition, 
the error from the local time evolution, and the truncation error.
Shown in Fig.~\ref{Heis_overlap_error} is the early-time error of the method 
analyzed using the Heisenberg model by following the error analysis in Zaletel \textit{et al.}\cite{Zaletel2015}
The N\'eel state is used for the initial state and the fidelity 
$\mathcal{F}(t) = 1-\braket{\Psi(t)|\Psi(t)_{\text{reference}}}$
is computed using 4\textsuperscript{th}-order iTEBD as a reference exact state.

We have used h-iTEBD with Krylov subspace dimension of 3 (h-iTEBD-K3) and 7 (h-iTEBD-K7). 
As expected, the early-time behavior differs significantly from the regime where the error from Krylov subspace expansion is
comparable to the subsequent Suzuki-Trotter step (h-iTEBD-K3), to that where the error is negligible (h-iTEBD-K7).
In the former regime, the error scales as $t^3$, 
which indicates that the most dominant error comes from the local time evolution (Krylov subspace expansion). 
In contrast, if the subspace dimension is sufficiently large (h-iTEBD-K7), 
the error scales as approximately $t^2$, as expected from the 2\textsuperscript{nd}-order Suzuki-Trotter decomposition.

The implementation of h-iTEBD is available as part of the Matrix Product Toolkit,\cite{mptoolkit} and further technical details on the method can be found in Ref.~\onlinecite{Hashizume_thesis}.

\begin{figure}[htb]
	\includegraphics[width=0.8\columnwidth]{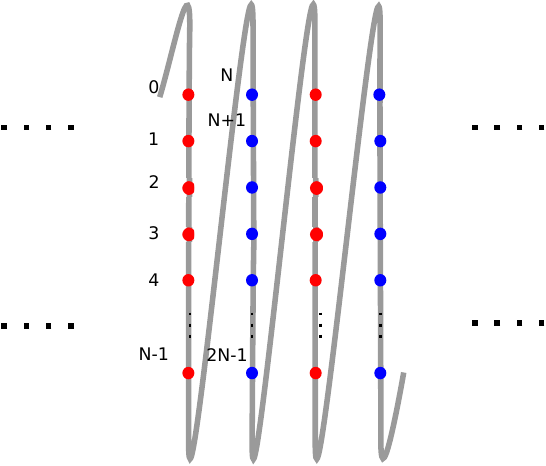}
	\caption{(Color online). A two-dimensional square lattice of size $N$ sites $\times$ $\infty$ sites and its mapping onto a one-dimensional chain with a unit cell of $N$ sites. The sites are numbered sequentially from top to bottom and then back to the top of the next vertical slice.
		\label{fig:mapping}}
\end{figure}

\begin{figure*}[htb]
\begin{minipage}{0.49\textwidth}
	\centering
	\includegraphics[width=\textwidth]{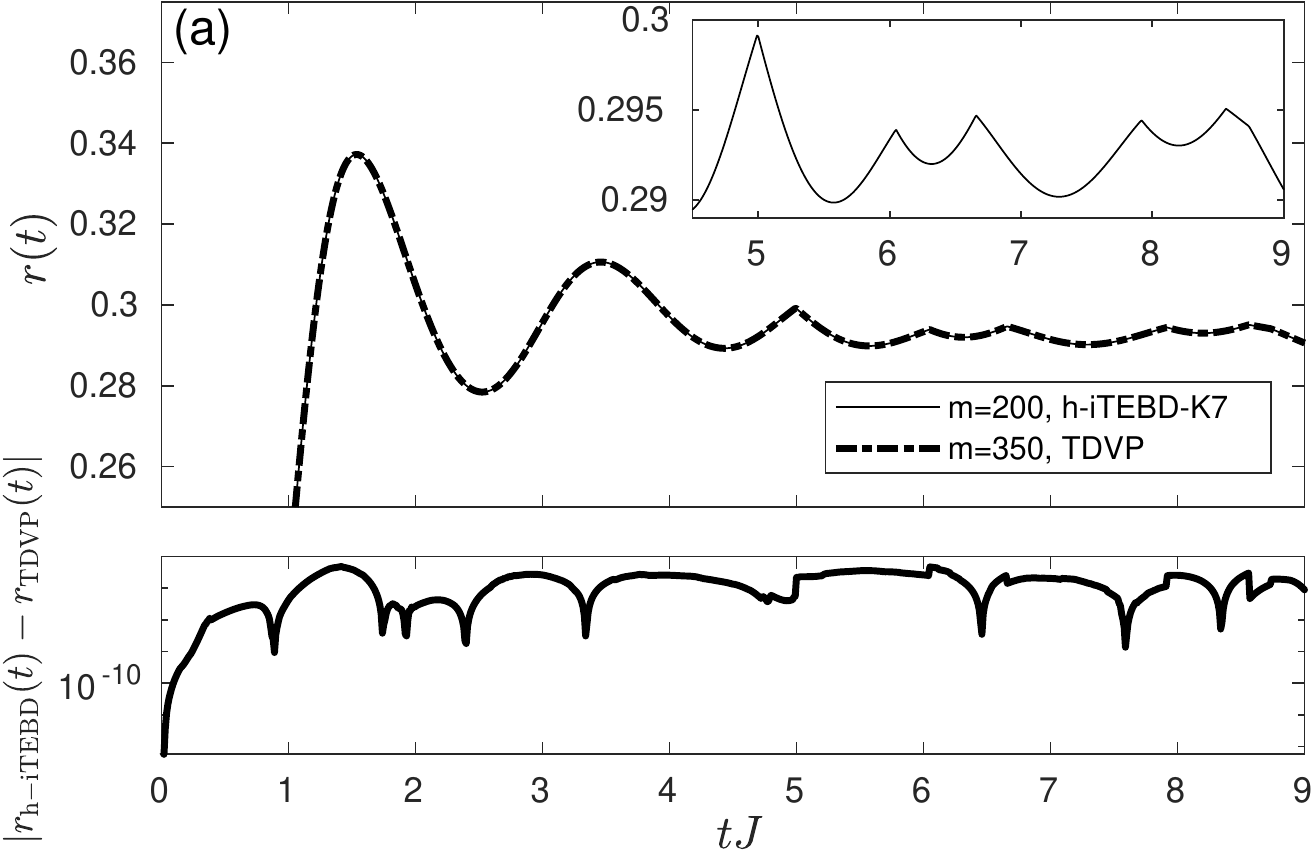}
\end{minipage}
\begin{minipage}{0.49\textwidth}
	\centering
	\includegraphics[width=\textwidth]{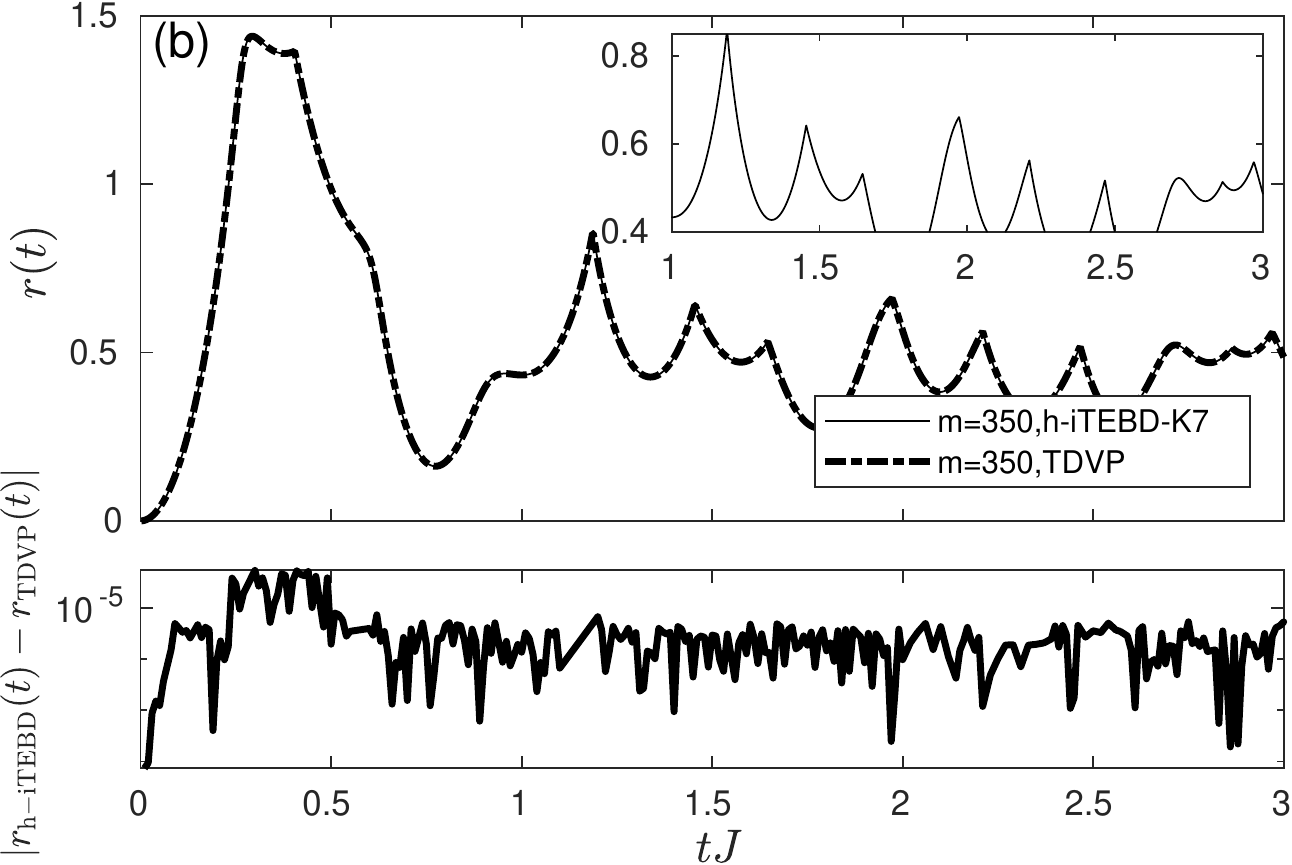}
\end{minipage}
\caption{
	Comparison of the Loschmidt-echo return rate calculated in h-iTEBD to that in TDVP for quenches in (a) the XXZ model from a N\'eel state, and 
	(b) the ANNNI model from a fully $z$-up polarized state (see text for details).
	The results show excellent agreement.}
\label{ANNNI_error}
\end{figure*}

\section{Mapping from two dimensions}\label{sec:mapping}
We now illustrate one of the major capabilities of our algorithm, which is the simulation of the dynamics of nonintegrable 2D quantum many-body models upon an appropriate mapping to a 1D chain with long-range fixed-length interactions that is translationally invariant down to a unit cell. We shall now describe this mapping.

For simplicity and without loss of generality, let us consider a square lattice of size $N$ sites $\times$ $\infty$ sites. 
If the lattice is translationally symmetric over the direction in which the length is infinite (horizontal according to Fig.~\ref{fig:mapping}),
then this lattice can be mapped onto an infinite one-dimensional chain with 
a unit cell of $N$ sites.

As shown in Fig.~\ref{fig:mapping}, 
the mapping can be done by numbering the sites along the direction in which the length is finite (vertical). 
Starting from site $0$ at the top, the site number increases going downwards. Upon reaching the bottom, we
continue on to the top of the next vertical slice to the right, and so on and so forth.
Obviously, sites $i$ and $i+N$ have the same interaction profile. The resulting 
one-dimensional chain is translationally symmetric down to a unit cell with $N$ sites, with a nontrivial long-range fixed-length interaction profile. Consequently, h-iTEBD can be used to simulate the time evolution on this type of lattice.

\section{Dynamical Quantum Phase Transitions}\label{sec:DQPT}

We also test our algorithm by calculating the Loschmidt-echo return rate\cite{Heyl2013}
\begin{align}\label{eq:RR}
    r(t)=-\lim_{L\to\infty}\frac{1}{L}\ln|\bra{\Psi_0}\exp(-\mathrm{i}\hat{H}t)\ket{\Psi_0}|^2,
\end{align}
where $\ket{\Psi_0}$ is an initial state quenched by a Hamiltonian $H$ at $t=0$.
This return rate is a dynamical analog of the thermal free energy in which complexified time $\mathrm{i}t$ stands for inverse temperature. Much the same way as nonanalyticities in the thermal free energy indicate critical temperatures at which thermal phase transitions occur, nonanalyticities in~\eqref{eq:RR} indicate \textit{critical times} at which \textit{dynamical quantum phase transitions} (DQPT) occur.\cite{Heyl2013,Heyl_review} The theory of DQPT has advanced significantly in the last few years, and experimental observations of the phenomenon have been possible in ion-trap\cite{Jurcevic2017} and ultracold-atom\cite{Flaeschner2018} setups.

Initially, it was demonstrated in the seminal work of Ref.~\onlinecite{Heyl2013} in the case of the nearest-neighbor transverse-field Ising chain that if a quench in the transverse-field strength crosses the equilibrium quantum critical point, cusps will appear in the return rate. However, quenching across the equilibrium quantum critical point was then shown to be in general neither a necessary nor a sufficient condition to obtain nonanalytic behavior in the return rate.\cite{Andraschko2014,Vajna2014} Indeed, the phenomenon of DQPT proved to be significantly richer with \textit{anomalous} cusps appearing for arbitrarily small quenches within the ordered phase\cite{Halimeh2017,Zauner2017,Homrighausen2017,Lang2017,Lang2018,Hashizume2018} when local excitations formed the lowest-lying quasiparticles in the spectrum of the quench Hamiltonian.\cite{Halimeh2018a,Defenu2019}

The study of DQPT in one spatial dimension has resulted in a very good understanding of the phenomenon in both integrable and nonintegrable systems. In higher dimensions the investigations have been mostly restricted to integrable models.\cite{Schmitt2015,Bhattacharya2017} However, we have recently used h-iTEBD to study DQPT in the nonintegrable two-dimensional transverse-field Ising model on a square lattice by mapping it onto a chain as explaied in Sec.~\ref{sec:mapping}.\cite{Hashizume2018} In addition to presenting DQPT results on this model on a triangular lattice, we also include in this Section benchmarking results of h-iTEBD in relation to classic examples from the field.

First, we present the benchmarking results in Fig.~\ref{ANNNI_error}, which
are direct comparisons of the Loschmidt-echo return rate obtained from h-iTEBD-K7 and TDVP for 
the nearest-neighbor XXZ chain in Fig.~\ref{ANNNI_error}(a), and the
axial next-nearest-neighbor Ising chain (ANNNI) in Fig.~\ref{ANNNI_error}(b).
The N\'eel state is quenched with the XXZ Hamiltonian 
\begin{align}
    \hat{H}_{\mathrm{XXZ}} (J_z) &= \sum_{l} \big(\hat{S}^{x}_{l}\hat{S}^{x}_{l+1} + \hat{S}^{y}_l \hat{S}^{y}_{l+1} 
    +J_z \hat{S}^{z}_{l}\hat{S}^{z}_{l+1}\big), \label{HNNXXZ}
\end{align}
at $J_z=1.2$, where $\hat{S}_i^{\{x,y,z\}}$ are the spin-$1/2$ operators on site $i$. We also quench the fully $z$-polarized state ($\Delta_i=0,h_i=0$) to $\Delta_f=-0.6,h_f=4.0$ in the ANNNI model given by the Hamiltonian
\begin{align}
    \hat{H}_{\mathrm{ANNNI}}(\Delta,h) &= 
    -J \sum_{l} \left[
        \hat{\sigma}^{z}_{l}\hat{\sigma}^{z}_{l+1} 
        +\Delta \hat{\sigma}^{z}_l\hat{\sigma}^{z}_{j+2} 
    +h\hat{\sigma}^{z}_{l} \right], \label{HANNI}
\end{align}
where $\hat{\sigma}_i^{\{x,y,z\}}$ are the Pauli operators on site $i$. For both quenches, the h-iTEBD results show great agreement with their TDVP counterparts. Furthermore, the ANNNI result shows full agreement with the result of Ref.~\onlinecite{Karrasch2013} for the same quench. 

\begin{figure}[htp]
	\centering
	\hspace{-.25 cm}\vspace{0.13cm}
	\includegraphics[width=.49\textwidth]{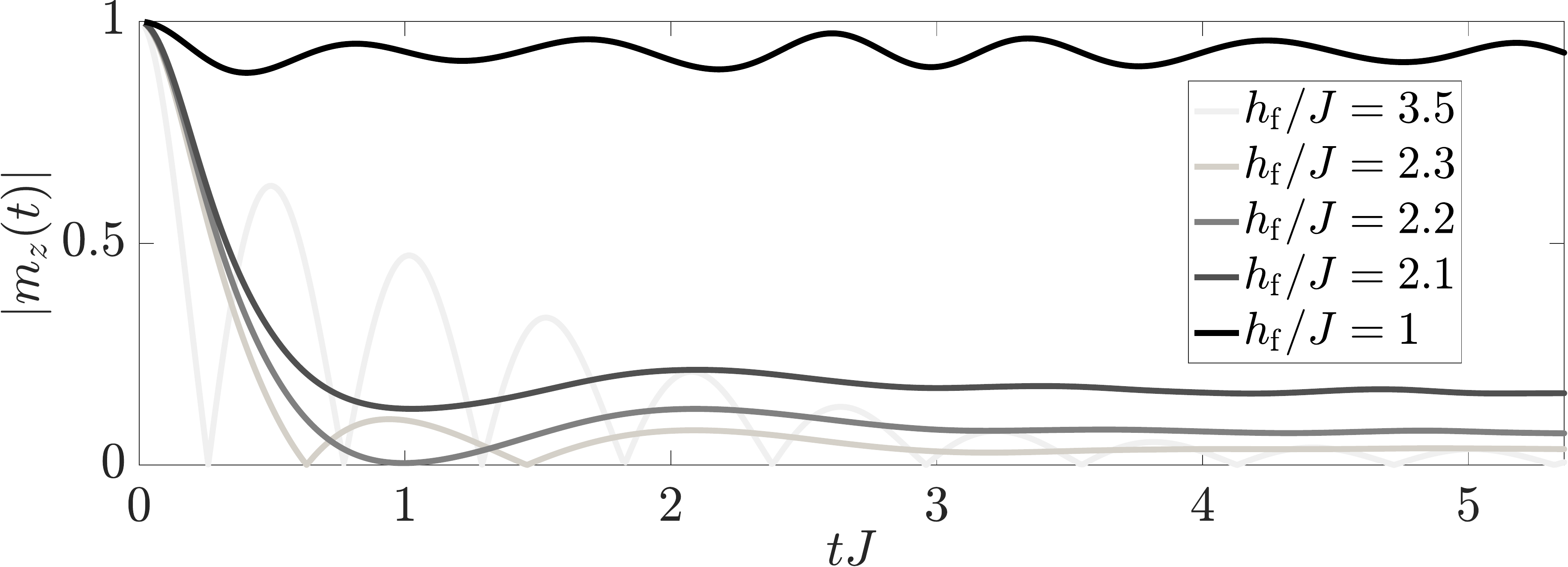}\\
	\hspace{-.25 cm}\vspace{0.13cm}
	\includegraphics[width=.49\textwidth]{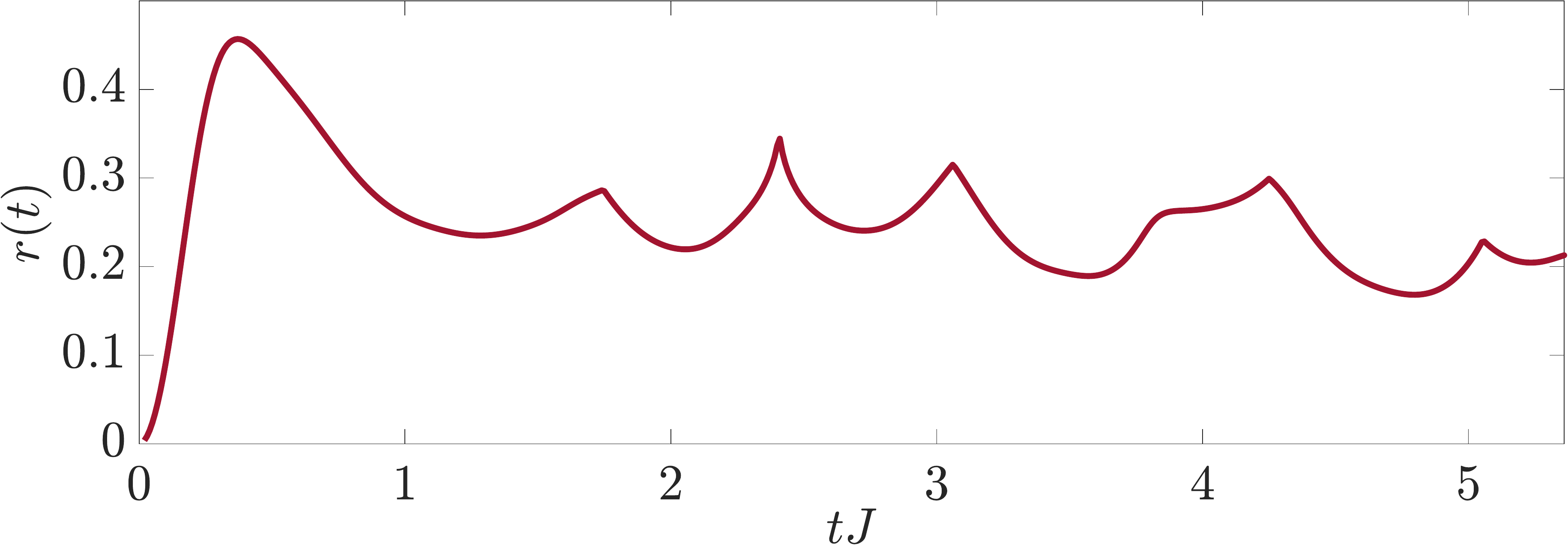}\\
	\caption{(Color online). Top panel shows the time evolution of the longitudinal magnetization in the ferromagnetic nearest-neighbor triangular lattice after a quench of the fully ordered state ($h_\text{i}=0$) with the Hamiltonian~\eqref{HTriangularIsing} at various final values $h_\text{f}$ of the transverse-field strength. The lower panel shows the return rate exhibiting anomalous cusps for a quench to $h_f=2.1J$, which is below the dynamical critical point $h_\text{c}^\text{d}\approx2.25J$.}
	\label{fig:mass} 
\end{figure}

In Fig.~\ref{fig:mass} we introduce new results on quench dynamics in the two-dimensional ferromagnetic quantum Ising model 
on a triangular lattice with the Hamiltonian
\begin{align}
    \hat{H}(h)=-J\sum_{\langle\mathbf{i},\mathbf{j}\rangle}
    \hat{\sigma}^z_\mathbf{i}\hat{\sigma}^z_\mathbf{j}-h\sum_\mathbf{j}\hat{\sigma}^x_\mathbf{j},
    \label{HTriangularIsing}
\end{align}
with $J>0$, which complement those of the same model on the square lattice in Ref.~\onlinecite{Hashizume2018}.
In both of these models, the system is implemented on an infinite cylinder. The Hamiltonian is constrained to have translation invariance, insomuch that each point on the lattice interacts with its six nearest neighbors. 
To use h-iTEBD, the model is projected onto an infinite one-dimensional lattice, as described in Sec.~\ref{sec:mapping}, that is translationally invariant down to a unit cell of size $N$ sites, where the 2D lattice is of size $N$ sites $\times$ $\infty$ sites. The projection breaks the two-dimensional translational invariance of the original Hamiltonian, but this does not remove the two-dimensional characteristics of the model. 
For example, the magnitude of the critical transverse-field strength where the equilibrium quantum phase transition occurs is $h_c^{e}\approx4.53J$ (see Appendix~\ref{AppendixECP}) as calculated iDMRG.
It is much larger than the equilibrium critical point of the one-dimensional model ($1J$) 
due to the higher connectivity which makes the system robust against fluctuations.
Thus the projected quasi-two-dimensional model has equilibrium characteristics beyond its one-dimensional counterpart.

Furthermore, as shown in Fig.~\ref{fig:mass}, the magnetization and the Loschmidt-echo return rate exhibit behavior that is fundamentally different from that in one-dimensional short-range models.
In particular, the top panel of Fig.~\ref{fig:mass} shows the dynamical critical point for quenches from $h_i=0$ to be $h_c^d\approx2.25J$, which separates between magnetization profiles that do not cross zero (for quenches to $h_f<h_c^d$) and those that do (quenches to $h_f>h_c^d$). We see that for small quenches below $h_c^d$ the magnetization does not show exponential decay but rather it relaxes to a finite nonzero value, characteristic of models with a finite-temperature phase transition. This can only be possible in spatial dimension $d>1$ for models with nearest-neighbor interactions and $\mathbb{Z}_2$ symmetry. Furthermore, we calculate the return rate in the wake of the quench from $h_i=0$ to $h_f=2.1J < h_c^d=2.25J$, shown in the bottom panel of Fig.~\ref{fig:mass}. The return rate shows clear \textit{anomalous}\cite{Halimeh2017} behavior in that the corresponding magnetization profile makes no zero crossings yet the return rate exhibits five nonanalytic cusps in the same time interval.\cite{Halimeh2018a} As discussed in detail in Refs.~\onlinecite{Hashizume2018} and~\onlinecite{Halimeh2018a}, these anomalous cusps are due to the lowest-lying excitations being local in nature. In the nearest-neighbor quantum Ising chain, this is never the case as the lowest excitations are freely propagating domain walls. As first shown in Ref.~\onlinecite{Halimeh2018a} and proven analytically in Ref.~\onlinecite{Defenu2019}, the presence of local excitations as the lowest-lying quasiparticles of the quench Hamiltonian is a necessary condition for anomalous cusps to arise. Hence, this clearly establishes that we are effectively working in 2D based on the equilirium and dynamical characteristics of this model.


\section{Conclusion}\label{sec:conc}
We have introduced an MPS algorithm for time-evolving systems with generic fixed-length long-range interactions.
The algorithm is a hybrid of a local time-evolution scheme and a global one.
An error analysis and a benchmarking of the algorithm has been carried out using Krylov-subspace expansion as the local time-evolution scheme and Suzuki-Trotter decomposition as the global one.

The analysis on scaling of the error with respect to the quasi-exact evolution result by 4\textsuperscript{th}-order iTEBD
shows behavior that agrees with the theoretical estimate. 
The benchmarking of the algorithm with respect to TDVP on integrable and nonintegrable models also shows excellent agreement.
It is shown that the algorithm is also applicable to short-ranged systems, as expected.

To go beyond local models, we applied the method to study DQPT in a quantum Ising triangular lattice placed on the two-dimensional 
surface of an infinite-length cylinder. 
The mapping of the lattice sites onto an infinite chain, gives rise to a Hamiltonian with nontrivial 
long-range interactions with translational symmetry down to a unit cell of size $N$ sites.

The simulated dynamics of the quantum Ising triangular model reveal, in both the order parameter and Loschmidt return rate, properties that are fundamentally different to dynamical behavior in short-range one-dimensional systems. In particular, we see that the magnetization in the wake of a small quench within the ferromagnetic phase decays to a finite nonzero value indicative of the presence of a finite-temperature phase transition. Indeed, such behavior cannot occur in short-range chains as the lower critical dimension for $\mathbb{Z}_2$ symmetry is $d=1$. Moreover, the return rate displays anomalous cusps, defined as cusps that occur without the order parameter ever making zero crossings, which is in stark contrast to the nearest-neighbor quantum Ising chain described in the seminal work of Ref.~\onlinecite{Heyl2013} on DQPT, where cusps are one-to-one connected to zeros of the magnetization.
The fact that the algorithm can calculate these critical phenomena in out-of-equilibrium illustrates its applicability and effectiveness in simulating the dynamics of models in dimensions $d>1$. It is worth mentioning that the results we have presented here on the quantum Ising triangular lattice are in full qualitative agreement with those of the quantum Ising square lattice of Ref.~\onlinecite{Hashizume2018}.

The future prospects lead in two directions.
Firstly, considering the use of different local time-evolution algorithms, such as TDVP, may yield much more precise results. This may actually resolve the necessity for truncating long-range interactions beyond the unit-cell length, and may allow the time evolution with truly long-range Hamiltonians. 
Secondly, this technique can be readily applied to many of the paradigmatic two-dimensional lattice models of current interest in state-of-the-art experiments in ion-trap and cold-atom laboratories. Determining the dynamical critical point, or at least calculating the short-time dynamics of such models can be fruitful on benchmarking and testing experimentally realizable strongly correlated systems.

\begin{acknowledgments}
T.H.~acknowledges useful discussions with Nariman Saadatmand.
J.C.H.~acknowledges support by the DFG Collaborative Research Centre SFB 1225 (ISO- QUANT) and the ERC Starting Grant StrEnQTh. I.P.M.~acknowledges support from the ARC Future Fellowships scheme, FT140100625.
\end{acknowledgments}
\bigskip
\bigskip
\appendix
\section{Equilibrium Quantum Critical Point of the Nearest-Neighbor Transverse-Field Ising Model on the Triangular Lattice
	\label{AppendixECP}
}

\begin{figure}[h!] 
\includegraphics[width=\columnwidth]{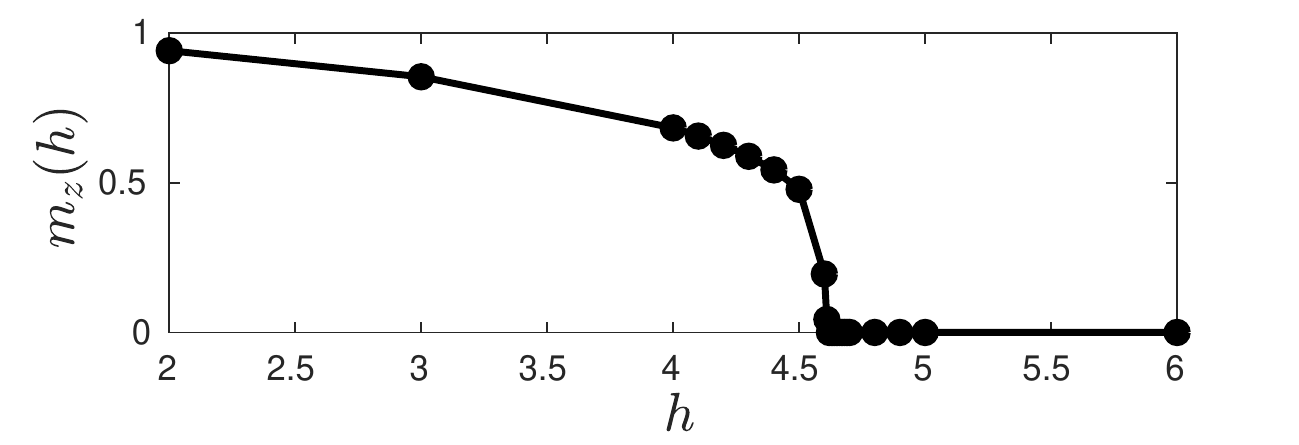}
\caption{Value of the longitudinal magnetization per site ($m_{z}=\sum_{i}\braket{\hat{\sigma}^{z}_i}/N$) with respect to the 
	transverse field strength $h$ of a triangular lattice of size $N=6$ sites $\times$ $\infty$ sites. 
	The equilibrium quantum phase transition is at $h\approx4.53J$, vastly different from its 1D counterpart due to much higher connectivity on the triangular lattice.\label{EQPdiagram}}
\end{figure}

Fig.~\ref{EQPdiagram} is the ground state of the quasi-two-dimensional ferromagnetic Ising model on a triangular lattice 
given by the Hamiltonian~\eqref{HTriangularIsing} with 6 sites per circumference. The phase diagram for its antiferromagnetic counterpart has been calculated in infinite DMRG (iDMRG) using a mapping similar to that in Sec.~\ref{sec:mapping} in Ref.~\onlinecite{Saadatmand2018}. 
An MPS ansatz of the ground state is calculated using iDMRG. 
The critical point of the quantum phase transition for lattice is predicted to happen at $h_{\mathrm{c}}\approx4.53J$.
This critical value is larger than that of the model on a square lattice and much larger than its 
one-dimensional counterpart.
This is due to the increased connectivity in the triangular lattice, which would then require a large critical field for fluctuations to be strong enough to break order in the system.

\bibliography{h-iTEBD_automatic.bib,pub.bib}

\begin{thebibliography}{69}%
\makeatletter
\providecommand \@ifxundefined [1]{%
 \@ifx{#1\undefined}
}%
\providecommand \@ifnum [1]{%
 \ifnum #1\expandafter \@firstoftwo
 \else \expandafter \@secondoftwo
 \fi
}%
\providecommand \@ifx [1]{%
 \ifx #1\expandafter \@firstoftwo
 \else \expandafter \@secondoftwo
 \fi
}%
\providecommand \natexlab [1]{#1}%
\providecommand \enquote  [1]{``#1''}%
\providecommand \bibnamefont  [1]{#1}%
\providecommand \bibfnamefont [1]{#1}%
\providecommand \citenamefont [1]{#1}%
\providecommand \href@noop [0]{\@secondoftwo}%
\providecommand \href [0]{\begingroup \@sanitize@url \@href}%
\providecommand \@href[1]{\@@startlink{#1}\@@href}%
\providecommand \@@href[1]{\endgroup#1\@@endlink}%
\providecommand \@sanitize@url [0]{\catcode `\\12\catcode `\$12\catcode
  `\&12\catcode `\#12\catcode `\^12\catcode `\_12\catcode `\%12\relax}%
\providecommand \@@startlink[1]{}%
\providecommand \@@endlink[0]{}%
\providecommand \url  [0]{\begingroup\@sanitize@url \@url }%
\providecommand \@url [1]{\endgroup\@href {#1}{\urlprefix }}%
\providecommand \urlprefix  [0]{URL }%
\providecommand \Eprint [0]{\href }%
\providecommand \doibase [0]{http://dx.doi.org/}%
\providecommand \selectlanguage [0]{\@gobble}%
\providecommand \bibinfo  [0]{\@secondoftwo}%
\providecommand \bibfield  [0]{\@secondoftwo}%
\providecommand \translation [1]{[#1]}%
\providecommand \BibitemOpen [0]{}%
\providecommand \bibitemStop [0]{}%
\providecommand \bibitemNoStop [0]{.\EOS\space}%
\providecommand \EOS [0]{\spacefactor3000\relax}%
\providecommand \BibitemShut  [1]{\csname bibitem#1\endcsname}%
\let\auto@bib@innerbib\@empty
\bibitem [{\citenamefont {Coleman}(2015)}]{Coleman_book}%
  \BibitemOpen
  \bibfield  {author} {\bibinfo {author} {\bibfnamefont {P.}~\bibnamefont
  {Coleman}},\ }\href {https://books.google.de/books?id=PDg1HQAACAAJ} {\emph
  {\bibinfo {title} {Introduction to Many-Body Physics}}}\ (\bibinfo
  {publisher} {Cambridge University Press},\ \bibinfo {year}
  {2015})\BibitemShut {NoStop}%
\bibitem [{\citenamefont {Kivelson}\ and\ \citenamefont
  {Kivelson}(2016)}]{Kivelson2016}%
  \BibitemOpen
  \bibfield  {author} {\bibinfo {author} {\bibfnamefont {S.}~\bibnamefont
  {Kivelson}}\ and\ \bibinfo {author} {\bibfnamefont {S.~A.}\ \bibnamefont
  {Kivelson}},\ }\href@noop {} {\bibfield  {journal} {\bibinfo  {journal} {npj
  Quantum Materials}\ }\textbf {\bibinfo {volume} {1}},\ \bibinfo {pages}
  {16024} (\bibinfo {year} {2016})}\BibitemShut {NoStop}%
\bibitem [{\citenamefont {Mori}\ \emph {et~al.}(2018)\citenamefont {Mori},
  \citenamefont {Ikeda}, \citenamefont {Kaminishi},\ and\ \citenamefont
  {Ueda}}]{Mori2018}%
  \BibitemOpen
  \bibfield  {author} {\bibinfo {author} {\bibfnamefont {T.}~\bibnamefont
  {Mori}}, \bibinfo {author} {\bibfnamefont {T.~N.}\ \bibnamefont {Ikeda}},
  \bibinfo {author} {\bibfnamefont {E.}~\bibnamefont {Kaminishi}}, \ and\
  \bibinfo {author} {\bibfnamefont {M.}~\bibnamefont {Ueda}},\ }\href
  {http://stacks.iop.org/0953-4075/51/i=11/a=112001} {\bibfield  {journal}
  {\bibinfo  {journal} {Journal of Physics B: Atomic, Molecular and Optical
  Physics}\ }\textbf {\bibinfo {volume} {51}},\ \bibinfo {pages} {112001}
  (\bibinfo {year} {2018})}\BibitemShut {NoStop}%
\bibitem [{\citenamefont {Zvyagin}(2016)}]{Zvyagin2016}%
  \BibitemOpen
  \bibfield  {author} {\bibinfo {author} {\bibfnamefont {A.~A.}\ \bibnamefont
  {Zvyagin}},\ }\href {\doibase 10.1063/1.4969869} {\bibfield  {journal}
  {\bibinfo  {journal} {Low Temperature Physics}\ }\textbf {\bibinfo {volume}
  {42}},\ \bibinfo {pages} {971} (\bibinfo {year} {2016})}\BibitemShut
  {NoStop}%
\bibitem [{\citenamefont {Sciolla}\ and\ \citenamefont
  {Biroli}(2010)}]{Sciolla2010}%
  \BibitemOpen
  \bibfield  {author} {\bibinfo {author} {\bibfnamefont {B.}~\bibnamefont
  {Sciolla}}\ and\ \bibinfo {author} {\bibfnamefont {G.}~\bibnamefont
  {Biroli}},\ }\href {\doibase 10.1103/PhysRevLett.105.220401} {\bibfield
  {journal} {\bibinfo  {journal} {Phys. Rev. Lett.}\ }\textbf {\bibinfo
  {volume} {105}},\ \bibinfo {pages} {220401} (\bibinfo {year}
  {2010})}\BibitemShut {NoStop}%
\bibitem [{\citenamefont {Sciolla}\ and\ \citenamefont
  {Biroli}(2011)}]{Sciolla2011}%
  \BibitemOpen
  \bibfield  {author} {\bibinfo {author} {\bibfnamefont {B.}~\bibnamefont
  {Sciolla}}\ and\ \bibinfo {author} {\bibfnamefont {G.}~\bibnamefont
  {Biroli}},\ }\href {http://stacks.iop.org/1742-5468/2011/i=11/a=P11003}
  {\bibfield  {journal} {\bibinfo  {journal} {Journal of Statistical Mechanics:
  Theory and Experiment}\ }\textbf {\bibinfo {volume} {2011}},\ \bibinfo
  {pages} {P11003} (\bibinfo {year} {2011})}\BibitemShut {NoStop}%
\bibitem [{\citenamefont {Calabrese}\ \emph {et~al.}(2011)\citenamefont
  {Calabrese}, \citenamefont {Essler},\ and\ \citenamefont
  {Fagotti}}]{Calabrese2011}%
  \BibitemOpen
  \bibfield  {author} {\bibinfo {author} {\bibfnamefont {P.}~\bibnamefont
  {Calabrese}}, \bibinfo {author} {\bibfnamefont {F.~H.~L.}\ \bibnamefont
  {Essler}}, \ and\ \bibinfo {author} {\bibfnamefont {M.}~\bibnamefont
  {Fagotti}},\ }\href {\doibase 10.1103/PhysRevLett.106.227203} {\bibfield
  {journal} {\bibinfo  {journal} {Phys. Rev. Lett.}\ }\textbf {\bibinfo
  {volume} {106}},\ \bibinfo {pages} {227203} (\bibinfo {year}
  {2011})}\BibitemShut {NoStop}%
\bibitem [{\citenamefont {Calabrese}\ \emph {et~al.}(2012)\citenamefont
  {Calabrese}, \citenamefont {Essler},\ and\ \citenamefont
  {Fagotti}}]{Calabrese2012}%
  \BibitemOpen
  \bibfield  {author} {\bibinfo {author} {\bibfnamefont {P.}~\bibnamefont
  {Calabrese}}, \bibinfo {author} {\bibfnamefont {F.~H.~L.}\ \bibnamefont
  {Essler}}, \ and\ \bibinfo {author} {\bibfnamefont {M.}~\bibnamefont
  {Fagotti}},\ }\href {\doibase 10.1088/1742-5468/2012/07/p07016} {\bibfield
  {journal} {\bibinfo  {journal} {Journal of Statistical Mechanics: Theory and
  Experiment}\ }\textbf {\bibinfo {volume} {2012}},\ \bibinfo {pages} {P07016}
  (\bibinfo {year} {2012})}\BibitemShut {NoStop}%
\bibitem [{\citenamefont {Altman}\ and\ \citenamefont
  {Auerbach}(2002)}]{Altman2002}%
  \BibitemOpen
  \bibfield  {author} {\bibinfo {author} {\bibfnamefont {E.}~\bibnamefont
  {Altman}}\ and\ \bibinfo {author} {\bibfnamefont {A.}~\bibnamefont
  {Auerbach}},\ }\href {\doibase 10.1103/PhysRevLett.89.250404} {\bibfield
  {journal} {\bibinfo  {journal} {Phys. Rev. Lett.}\ }\textbf {\bibinfo
  {volume} {89}},\ \bibinfo {pages} {250404} (\bibinfo {year}
  {2002})}\BibitemShut {NoStop}%
\bibitem [{\citenamefont {Barmettler}\ \emph {et~al.}(2009)\citenamefont
  {Barmettler}, \citenamefont {Punk}, \citenamefont {Gritsev}, \citenamefont
  {Demler},\ and\ \citenamefont {Altman}}]{Barmettler2009}%
  \BibitemOpen
  \bibfield  {author} {\bibinfo {author} {\bibfnamefont {P.}~\bibnamefont
  {Barmettler}}, \bibinfo {author} {\bibfnamefont {M.}~\bibnamefont {Punk}},
  \bibinfo {author} {\bibfnamefont {V.}~\bibnamefont {Gritsev}}, \bibinfo
  {author} {\bibfnamefont {E.}~\bibnamefont {Demler}}, \ and\ \bibinfo {author}
  {\bibfnamefont {E.}~\bibnamefont {Altman}},\ }\href {\doibase
  10.1103/PhysRevLett.102.130603} {\bibfield  {journal} {\bibinfo  {journal}
  {Phys. Rev. Lett.}\ }\textbf {\bibinfo {volume} {102}},\ \bibinfo {pages}
  {130603} (\bibinfo {year} {2009})}\BibitemShut {NoStop}%
\bibitem [{\citenamefont {Halimeh}\ \emph {et~al.}(2017)\citenamefont
  {Halimeh}, \citenamefont {Zauner-Stauber}, \citenamefont {McCulloch},
  \citenamefont {de~Vega}, \citenamefont {Schollw\"ock},\ and\ \citenamefont
  {Kastner}}]{Halimeh2017b}%
  \BibitemOpen
  \bibfield  {author} {\bibinfo {author} {\bibfnamefont {J.~C.}\ \bibnamefont
  {Halimeh}}, \bibinfo {author} {\bibfnamefont {V.}~\bibnamefont
  {Zauner-Stauber}}, \bibinfo {author} {\bibfnamefont {I.~P.}\ \bibnamefont
  {McCulloch}}, \bibinfo {author} {\bibfnamefont {I.}~\bibnamefont {de~Vega}},
  \bibinfo {author} {\bibfnamefont {U.}~\bibnamefont {Schollw\"ock}}, \ and\
  \bibinfo {author} {\bibfnamefont {M.}~\bibnamefont {Kastner}},\ }\href
  {\doibase 10.1103/PhysRevB.95.024302} {\bibfield  {journal} {\bibinfo
  {journal} {Phys. Rev. B}\ }\textbf {\bibinfo {volume} {95}},\ \bibinfo
  {pages} {024302} (\bibinfo {year} {2017})}\BibitemShut {NoStop}%
\bibitem [{\citenamefont {Heyl}\ \emph {et~al.}(2013)\citenamefont {Heyl},
  \citenamefont {Polkovnikov},\ and\ \citenamefont {Kehrein}}]{Heyl2013}%
  \BibitemOpen
  \bibfield  {author} {\bibinfo {author} {\bibfnamefont {M.}~\bibnamefont
  {Heyl}}, \bibinfo {author} {\bibfnamefont {A.}~\bibnamefont {Polkovnikov}}, \
  and\ \bibinfo {author} {\bibfnamefont {S.}~\bibnamefont {Kehrein}},\ }\href
  {\doibase 10.1103/PhysRevLett.110.135704} {\bibfield  {journal} {\bibinfo
  {journal} {Physical Review Letters}\ }\textbf {\bibinfo {volume} {110}},\
  \bibinfo {pages} {135704} (\bibinfo {year} {2013})},\ \Eprint
  {http://arxiv.org/abs/1206.2505} {arXiv:1206.2505} \BibitemShut {NoStop}%
\bibitem [{\citenamefont {Heyl}(2018)}]{Heyl_review}%
  \BibitemOpen
  \bibfield  {author} {\bibinfo {author} {\bibfnamefont {M.}~\bibnamefont
  {Heyl}},\ }\href {http://stacks.iop.org/0034-4885/81/i=5/a=054001} {\bibfield
   {journal} {\bibinfo  {journal} {Reports on Progress in Physics}\ }\textbf
  {\bibinfo {volume} {81}},\ \bibinfo {pages} {054001} (\bibinfo {year}
  {2018})}\BibitemShut {NoStop}%
\bibitem [{\citenamefont {White}(1992)}]{White1992}%
  \BibitemOpen
  \bibfield  {author} {\bibinfo {author} {\bibfnamefont {S.~R.}\ \bibnamefont
  {White}},\ }\href {\doibase 10.1103/PhysRevLett.69.2863} {\bibfield
  {journal} {\bibinfo  {journal} {Physical Review Letters}\ }\textbf {\bibinfo
  {volume} {69}},\ \bibinfo {pages} {2863} (\bibinfo {year}
  {1992})}\BibitemShut {NoStop}%
\bibitem [{\citenamefont {Schollw{\"{o}}ck}(2005)}]{Schollwock2005a}%
  \BibitemOpen
  \bibfield  {author} {\bibinfo {author} {\bibfnamefont {U.}~\bibnamefont
  {Schollw{\"{o}}ck}},\ }\href {\doibase 10.1103/RevModPhys.77.259} {\bibfield
  {journal} {\bibinfo  {journal} {Reviews of Modern Physics}\ }\textbf
  {\bibinfo {volume} {77}},\ \bibinfo {pages} {259} (\bibinfo {year} {2005})},\
  \Eprint {http://arxiv.org/abs/0409292v1} {arXiv:0409292v1 [arXiv:cond-mat]}
  \BibitemShut {NoStop}%
\bibitem [{\citenamefont {Schollwöck}(2011)}]{Uli2011}%
  \BibitemOpen
  \bibfield  {author} {\bibinfo {author} {\bibfnamefont {U.}~\bibnamefont
  {Schollwöck}},\ }\href {\doibase https://doi.org/10.1016/j.aop.2010.09.012}
  {\bibfield  {journal} {\bibinfo  {journal} {Annals of Physics}\ }\textbf
  {\bibinfo {volume} {326}},\ \bibinfo {pages} {96 } (\bibinfo {year}
  {2011})},\ \bibinfo {note} {january 2011 Special Issue}\BibitemShut {NoStop}%
\bibitem [{\citenamefont {Vidal}(2003)}]{Vidal2003}%
  \BibitemOpen
  \bibfield  {author} {\bibinfo {author} {\bibfnamefont {G.}~\bibnamefont
  {Vidal}},\ }\href {\doibase 10.1103/PhysRevLett.91.147902} {\bibfield
  {journal} {\bibinfo  {journal} {Physical Review Letters}\ }\textbf {\bibinfo
  {volume} {91}},\ \bibinfo {pages} {147902} (\bibinfo {year} {2003})},\
  \Eprint {http://arxiv.org/abs/quant-ph/0301063} {arXiv:quant-ph/0301063
  [quant-ph]} \BibitemShut {NoStop}%
\bibitem [{\citenamefont {Daley}\ \emph {et~al.}(2004)\citenamefont {Daley},
  \citenamefont {Kollath}, \citenamefont {Schollw{\"{o}}ck},\ and\
  \citenamefont {Vidal}}]{Daley2004}%
  \BibitemOpen
  \bibfield  {author} {\bibinfo {author} {\bibfnamefont {A.~J.}\ \bibnamefont
  {Daley}}, \bibinfo {author} {\bibfnamefont {C.}~\bibnamefont {Kollath}},
  \bibinfo {author} {\bibfnamefont {U.}~\bibnamefont {Schollw{\"{o}}ck}}, \
  and\ \bibinfo {author} {\bibfnamefont {G.}~\bibnamefont {Vidal}},\ }\href
  {\doibase 10.1088/1742-5468/2004/04/P04005} {\bibfield  {journal} {\bibinfo
  {journal} {Journal of Statistical Mechanics: Theory and Experiment}\ }\textbf
  {\bibinfo {volume} {2004}},\ \bibinfo {pages} {P04005} (\bibinfo {year}
  {2004})},\ \Eprint {http://arxiv.org/abs/cond-mat/0403313}
  {arXiv:cond-mat/0403313 [cond-mat]} \BibitemShut {NoStop}%
\bibitem [{\citenamefont {Haegeman}\ \emph {et~al.}(2011)\citenamefont
  {Haegeman}, \citenamefont {Cirac}, \citenamefont {Osborne}, \citenamefont
  {Pi{\v{z}}orn}, \citenamefont {Verschelde},\ and\ \citenamefont
  {Verstraete}}]{Haegeman2011}%
  \BibitemOpen
  \bibfield  {author} {\bibinfo {author} {\bibfnamefont {J.}~\bibnamefont
  {Haegeman}}, \bibinfo {author} {\bibfnamefont {J.~I.}\ \bibnamefont {Cirac}},
  \bibinfo {author} {\bibfnamefont {T.~J.}\ \bibnamefont {Osborne}}, \bibinfo
  {author} {\bibfnamefont {I.}~\bibnamefont {Pi{\v{z}}orn}}, \bibinfo {author}
  {\bibfnamefont {H.}~\bibnamefont {Verschelde}}, \ and\ \bibinfo {author}
  {\bibfnamefont {F.}~\bibnamefont {Verstraete}},\ }\href {\doibase
  10.1103/PhysRevLett.107.070601} {\bibfield  {journal} {\bibinfo  {journal}
  {Physical Review Letters}\ }\textbf {\bibinfo {volume} {107}},\ \bibinfo
  {pages} {070601} (\bibinfo {year} {2011})},\ \Eprint
  {http://arxiv.org/abs/1103.0936} {arXiv:1103.0936} \BibitemShut {NoStop}%
\bibitem [{\citenamefont {Garc{\'{i}}a-Ripoll}(2006)}]{GarcIa-Ripoll2006}%
  \BibitemOpen
  \bibfield  {author} {\bibinfo {author} {\bibfnamefont {J.~J.}\ \bibnamefont
  {Garc{\'{i}}a-Ripoll}},\ }\href {\doibase 10.1088/1367-2630/8/12/305}
  {\bibfield  {journal} {\bibinfo  {journal} {New Journal of Physics}\ }\textbf
  {\bibinfo {volume} {8}},\ \bibinfo {pages} {305} (\bibinfo {year} {2006})},\
  \Eprint {http://arxiv.org/abs/0610210} {arXiv:0610210 [cond-mat]}
  \BibitemShut {NoStop}%
\bibitem [{\citenamefont {Zaletel}\ \emph {et~al.}(2015)\citenamefont
  {Zaletel}, \citenamefont {Mong}, \citenamefont {Karrasch}, \citenamefont
  {Moore},\ and\ \citenamefont {Pollmann}}]{Zaletel2015}%
  \BibitemOpen
  \bibfield  {author} {\bibinfo {author} {\bibfnamefont {M.~P.}\ \bibnamefont
  {Zaletel}}, \bibinfo {author} {\bibfnamefont {R.~S.~K.}\ \bibnamefont
  {Mong}}, \bibinfo {author} {\bibfnamefont {C.}~\bibnamefont {Karrasch}},
  \bibinfo {author} {\bibfnamefont {J.~E.}\ \bibnamefont {Moore}}, \ and\
  \bibinfo {author} {\bibfnamefont {F.}~\bibnamefont {Pollmann}},\ }\href
  {\doibase 10.1103/PhysRevB.91.165112} {\bibfield  {journal} {\bibinfo
  {journal} {Physical Review B}\ }\textbf {\bibinfo {volume} {91}},\ \bibinfo
  {pages} {165112} (\bibinfo {year} {2015})},\ \Eprint
  {http://arxiv.org/abs/1407.1832} {arXiv:1407.1832} \BibitemShut {NoStop}%
\bibitem [{\citenamefont {Jurcevic}\ \emph {et~al.}(2017)\citenamefont
  {Jurcevic}, \citenamefont {Shen}, \citenamefont {Hauke}, \citenamefont
  {Maier}, \citenamefont {Brydges}, \citenamefont {Hempel}, \citenamefont
  {Lanyon}, \citenamefont {Heyl}, \citenamefont {Blatt},\ and\ \citenamefont
  {Roos}}]{Jurcevic2017}%
  \BibitemOpen
  \bibfield  {author} {\bibinfo {author} {\bibfnamefont {P.}~\bibnamefont
  {Jurcevic}}, \bibinfo {author} {\bibfnamefont {H.}~\bibnamefont {Shen}},
  \bibinfo {author} {\bibfnamefont {P.}~\bibnamefont {Hauke}}, \bibinfo
  {author} {\bibfnamefont {C.}~\bibnamefont {Maier}}, \bibinfo {author}
  {\bibfnamefont {T.}~\bibnamefont {Brydges}}, \bibinfo {author} {\bibfnamefont
  {C.}~\bibnamefont {Hempel}}, \bibinfo {author} {\bibfnamefont {B.~P.}\
  \bibnamefont {Lanyon}}, \bibinfo {author} {\bibfnamefont {M.}~\bibnamefont
  {Heyl}}, \bibinfo {author} {\bibfnamefont {R.}~\bibnamefont {Blatt}}, \ and\
  \bibinfo {author} {\bibfnamefont {C.~F.}\ \bibnamefont {Roos}},\ }\href
  {\doibase 10.1103/PhysRevLett.119.080501} {\bibfield  {journal} {\bibinfo
  {journal} {Phys. Rev. Lett.}\ }\textbf {\bibinfo {volume} {119}},\ \bibinfo
  {pages} {080501} (\bibinfo {year} {2017})}\BibitemShut {NoStop}%
\bibitem [{\citenamefont {Fl{\"a}schner}\ \emph {et~al.}(2018)\citenamefont
  {Fl{\"a}schner}, \citenamefont {Vogel}, \citenamefont {Tarnowski},
  \citenamefont {Rem}, \citenamefont {L{\"u}hmann}, \citenamefont {Heyl},
  \citenamefont {Budich}, \citenamefont {Mathey}, \citenamefont {Sengstock},\
  and\ \citenamefont {Weitenberg}}]{Flaeschner2018}%
  \BibitemOpen
  \bibfield  {author} {\bibinfo {author} {\bibfnamefont {N.}~\bibnamefont
  {Fl{\"a}schner}}, \bibinfo {author} {\bibfnamefont {D.}~\bibnamefont
  {Vogel}}, \bibinfo {author} {\bibfnamefont {M.}~\bibnamefont {Tarnowski}},
  \bibinfo {author} {\bibfnamefont {B.~S.}\ \bibnamefont {Rem}}, \bibinfo
  {author} {\bibfnamefont {D.-S.}\ \bibnamefont {L{\"u}hmann}}, \bibinfo
  {author} {\bibfnamefont {M.}~\bibnamefont {Heyl}}, \bibinfo {author}
  {\bibfnamefont {J.~C.}\ \bibnamefont {Budich}}, \bibinfo {author}
  {\bibfnamefont {L.}~\bibnamefont {Mathey}}, \bibinfo {author} {\bibfnamefont
  {K.}~\bibnamefont {Sengstock}}, \ and\ \bibinfo {author} {\bibfnamefont
  {C.}~\bibnamefont {Weitenberg}},\ }\href
  {https://doi.org/10.1038/s41567-017-0013-8} {\bibfield  {journal} {\bibinfo
  {journal} {Nature Physics}\ }\textbf {\bibinfo {volume} {14}},\ \bibinfo
  {pages} {265} (\bibinfo {year} {2018})}\BibitemShut {NoStop}%
\bibitem [{\citenamefont {McCulloch}(2008{\natexlab{a}})}]{Bloch2008a}%
  \BibitemOpen
  \bibfield  {author} {\bibinfo {author} {\bibfnamefont {I.~P.}\ \bibnamefont
  {McCulloch}},\ }\href {\doibase 10.1103/RevModPhys.80.885} {\bibfield
  {journal} {\bibinfo  {journal} {Reviews of Modern Physics}\ }\textbf
  {\bibinfo {volume} {80}},\ \bibinfo {pages} {885} (\bibinfo {year}
  {2008}{\natexlab{a}})},\ \Eprint {http://arxiv.org/abs/0804.2509}
  {arXiv:0804.2509} \BibitemShut {NoStop}%
\bibitem [{\citenamefont {Britton}\ \emph {et~al.}(2012)\citenamefont
  {Britton}, \citenamefont {Sawyer}, \citenamefont {Keith}, \citenamefont
  {Wang}, \citenamefont {Freericks}, \citenamefont {Uys}, \citenamefont
  {Biercuk},\ and\ \citenamefont {Bollinger}}]{Britton2012}%
  \BibitemOpen
  \bibfield  {author} {\bibinfo {author} {\bibfnamefont {J.~W.}\ \bibnamefont
  {Britton}}, \bibinfo {author} {\bibfnamefont {B.~C.}\ \bibnamefont {Sawyer}},
  \bibinfo {author} {\bibfnamefont {A.~C.}\ \bibnamefont {Keith}}, \bibinfo
  {author} {\bibfnamefont {C.-C.~J.}\ \bibnamefont {Wang}}, \bibinfo {author}
  {\bibfnamefont {J.~K.}\ \bibnamefont {Freericks}}, \bibinfo {author}
  {\bibfnamefont {H.}~\bibnamefont {Uys}}, \bibinfo {author} {\bibfnamefont
  {M.~J.}\ \bibnamefont {Biercuk}}, \ and\ \bibinfo {author} {\bibfnamefont
  {J.~J.}\ \bibnamefont {Bollinger}},\ }\href {\doibase 10.1038/nature10981}
  {\bibfield  {journal} {\bibinfo  {journal} {Nature}\ }\textbf {\bibinfo
  {volume} {484}},\ \bibinfo {pages} {489} (\bibinfo {year} {2012})},\ \Eprint
  {http://arxiv.org/abs/1204.5789} {arXiv:1204.5789} \BibitemShut {NoStop}%
\bibitem [{\citenamefont {Georgescu}\ \emph {et~al.}(2014)\citenamefont
  {Georgescu}, \citenamefont {Ashhab},\ and\ \citenamefont
  {Nori}}]{Georgescu2014}%
  \BibitemOpen
  \bibfield  {author} {\bibinfo {author} {\bibfnamefont {I.~M.}\ \bibnamefont
  {Georgescu}}, \bibinfo {author} {\bibfnamefont {S.}~\bibnamefont {Ashhab}}, \
  and\ \bibinfo {author} {\bibfnamefont {F.}~\bibnamefont {Nori}},\ }\href
  {\doibase 10.1103/RevModPhys.86.153} {\bibfield  {journal} {\bibinfo
  {journal} {Reviews of Modern Physics}\ }\textbf {\bibinfo {volume} {86}},\
  \bibinfo {pages} {153} (\bibinfo {year} {2014})}\BibitemShut {NoStop}%
\bibitem [{\citenamefont {Hung}\ \emph {et~al.}(2016)\citenamefont {Hung},
  \citenamefont {Gonz{\'{a}}lez-Tudela}, \citenamefont {Cirac},\ and\
  \citenamefont {Kimble}}]{Hung2016}%
  \BibitemOpen
  \bibfield  {author} {\bibinfo {author} {\bibfnamefont {C.-L.}\ \bibnamefont
  {Hung}}, \bibinfo {author} {\bibfnamefont {A.}~\bibnamefont
  {Gonz{\'{a}}lez-Tudela}}, \bibinfo {author} {\bibfnamefont {J.~I.}\
  \bibnamefont {Cirac}}, \ and\ \bibinfo {author} {\bibfnamefont {H.~J.}\
  \bibnamefont {Kimble}},\ }\href {\doibase 10.1073/pnas.1603777113} {\bibfield
   {journal} {\bibinfo  {journal} {Proceedings of the National Academy of
  Sciences}\ }\textbf {\bibinfo {volume} {113}},\ \bibinfo {pages} {E4946}
  (\bibinfo {year} {2016})},\ \Eprint {http://arxiv.org/abs/1603.05860}
  {arXiv:1603.05860} \BibitemShut {NoStop}%
\bibitem [{\citenamefont {Bentsen}\ \emph {et~al.}(2019)\citenamefont
  {Bentsen}, \citenamefont {Hashizume}, \citenamefont {Buyskikh}, \citenamefont
  {Davis}, \citenamefont {Daley}, \citenamefont {Gubser},\ and\ \citenamefont
  {Schleier-Smith}}]{Bentsen2019}%
  \BibitemOpen
  \bibfield  {author} {\bibinfo {author} {\bibfnamefont {G.}~\bibnamefont
  {Bentsen}}, \bibinfo {author} {\bibfnamefont {T.}~\bibnamefont {Hashizume}},
  \bibinfo {author} {\bibfnamefont {A.~S.}\ \bibnamefont {Buyskikh}}, \bibinfo
  {author} {\bibfnamefont {E.~J.}\ \bibnamefont {Davis}}, \bibinfo {author}
  {\bibfnamefont {A.~J.}\ \bibnamefont {Daley}}, \bibinfo {author}
  {\bibfnamefont {S.~S.}\ \bibnamefont {Gubser}}, \ and\ \bibinfo {author}
  {\bibfnamefont {M.}~\bibnamefont {Schleier-Smith}},\ }\href {\doibase
  10.1103/PhysRevLett.123.130601} {\bibfield  {journal} {\bibinfo  {journal}
  {Physical Review Letters}\ }\textbf {\bibinfo {volume} {123}},\ \bibinfo
  {pages} {130601} (\bibinfo {year} {2019})},\ \Eprint
  {http://arxiv.org/abs/1905.11430} {arXiv:1905.11430} \BibitemShut {NoStop}%
\bibitem [{\citenamefont {Martinez}\ \emph {et~al.}(2016)\citenamefont
  {Martinez}, \citenamefont {Muschik}, \citenamefont {Schindler}, \citenamefont
  {Nigg}, \citenamefont {Erhard}, \citenamefont {Heyl}, \citenamefont {Hauke},
  \citenamefont {Dalmonte}, \citenamefont {Monz}, \citenamefont {Zoller},\ and\
  \citenamefont {Blatt}}]{Martinez2016}%
  \BibitemOpen
  \bibfield  {author} {\bibinfo {author} {\bibfnamefont {E.~A.}\ \bibnamefont
  {Martinez}}, \bibinfo {author} {\bibfnamefont {C.~A.}\ \bibnamefont
  {Muschik}}, \bibinfo {author} {\bibfnamefont {P.}~\bibnamefont {Schindler}},
  \bibinfo {author} {\bibfnamefont {D.}~\bibnamefont {Nigg}}, \bibinfo {author}
  {\bibfnamefont {A.}~\bibnamefont {Erhard}}, \bibinfo {author} {\bibfnamefont
  {M.}~\bibnamefont {Heyl}}, \bibinfo {author} {\bibfnamefont {P.}~\bibnamefont
  {Hauke}}, \bibinfo {author} {\bibfnamefont {M.}~\bibnamefont {Dalmonte}},
  \bibinfo {author} {\bibfnamefont {T.}~\bibnamefont {Monz}}, \bibinfo {author}
  {\bibfnamefont {P.}~\bibnamefont {Zoller}}, \ and\ \bibinfo {author}
  {\bibfnamefont {R.}~\bibnamefont {Blatt}},\ }\href
  {https://doi.org/10.1038/nature18318} {\bibfield  {journal} {\bibinfo
  {journal} {Nature}\ }\textbf {\bibinfo {volume} {534}},\ \bibinfo {pages}
  {516 EP } (\bibinfo {year} {2016})}\BibitemShut {NoStop}%
\bibitem [{\citenamefont {Schweizer}\ \emph {et~al.}(2019)\citenamefont
  {Schweizer}, \citenamefont {Grusdt}, \citenamefont {Berngruber},
  \citenamefont {Barbiero}, \citenamefont {Demler}, \citenamefont {Goldman},
  \citenamefont {Bloch},\ and\ \citenamefont {Aidelsburger}}]{Schweizer2019}%
  \BibitemOpen
  \bibfield  {author} {\bibinfo {author} {\bibfnamefont {C.}~\bibnamefont
  {Schweizer}}, \bibinfo {author} {\bibfnamefont {F.}~\bibnamefont {Grusdt}},
  \bibinfo {author} {\bibfnamefont {M.}~\bibnamefont {Berngruber}}, \bibinfo
  {author} {\bibfnamefont {L.}~\bibnamefont {Barbiero}}, \bibinfo {author}
  {\bibfnamefont {E.}~\bibnamefont {Demler}}, \bibinfo {author} {\bibfnamefont
  {N.}~\bibnamefont {Goldman}}, \bibinfo {author} {\bibfnamefont
  {I.}~\bibnamefont {Bloch}}, \ and\ \bibinfo {author} {\bibfnamefont
  {M.}~\bibnamefont {Aidelsburger}},\ }\href {\doibase
  10.1038/s41567-019-0649-7} {\bibfield  {journal} {\bibinfo  {journal} {Nature
  Physics}\ } (\bibinfo {year} {2019}),\ 10.1038/s41567-019-0649-7}\BibitemShut
  {NoStop}%
\bibitem [{\citenamefont {{Mil}}\ \emph {et~al.}(2019)\citenamefont {{Mil}},
  \citenamefont {{Zache}}, \citenamefont {{Hegde}}, \citenamefont {{Xia}},
  \citenamefont {{Bhatt}}, \citenamefont {{Oberthaler}}, \citenamefont
  {{Hauke}}, \citenamefont {{Berges}},\ and\ \citenamefont
  {{Jendrzejewski}}}]{Mil2019}%
  \BibitemOpen
  \bibfield  {author} {\bibinfo {author} {\bibfnamefont {A.}~\bibnamefont
  {{Mil}}}, \bibinfo {author} {\bibfnamefont {T.~V.}\ \bibnamefont {{Zache}}},
  \bibinfo {author} {\bibfnamefont {A.}~\bibnamefont {{Hegde}}}, \bibinfo
  {author} {\bibfnamefont {A.}~\bibnamefont {{Xia}}}, \bibinfo {author}
  {\bibfnamefont {R.~P.}\ \bibnamefont {{Bhatt}}}, \bibinfo {author}
  {\bibfnamefont {M.~K.}\ \bibnamefont {{Oberthaler}}}, \bibinfo {author}
  {\bibfnamefont {P.}~\bibnamefont {{Hauke}}}, \bibinfo {author} {\bibfnamefont
  {J.}~\bibnamefont {{Berges}}}, \ and\ \bibinfo {author} {\bibfnamefont
  {F.}~\bibnamefont {{Jendrzejewski}}},\ }\href
  {https://arxiv.org/abs/1909.07641} {\bibfield  {journal} {\bibinfo  {journal}
  {ArXiv e-prints}\ } (\bibinfo {year} {2019})},\ \Eprint
  {http://arxiv.org/abs/1909.07641} {arXiv:1909.07641 [cond-mat.quant-gas]}
  \BibitemShut {NoStop}%
\bibitem [{\citenamefont {Arute}\ \emph {et~al.}(2019)\citenamefont {Arute},
  \citenamefont {Arya}, \citenamefont {Babbush}, \citenamefont {Bacon},
  \citenamefont {Bardin}, \citenamefont {Barends}, \citenamefont {Biswas},
  \citenamefont {Boixo}, \citenamefont {Brandao}, \citenamefont {Buell},
  \citenamefont {Burkett}, \citenamefont {Chen}, \citenamefont {Chen},
  \citenamefont {Chiaro}, \citenamefont {Collins}, \citenamefont {Courtney},
  \citenamefont {Dunsworth}, \citenamefont {Farhi}, \citenamefont {Foxen},
  \citenamefont {Fowler}, \citenamefont {Gidney}, \citenamefont {Giustina},
  \citenamefont {Graff}, \citenamefont {Guerin}, \citenamefont {Habegger},
  \citenamefont {Harrigan}, \citenamefont {Hartmann}, \citenamefont {Ho},
  \citenamefont {Hoffmann}, \citenamefont {Huang}, \citenamefont {Humble},
  \citenamefont {Isakov}, \citenamefont {Jeffrey}, \citenamefont {Jiang},
  \citenamefont {Kafri}, \citenamefont {Kechedzhi}, \citenamefont {Kelly},
  \citenamefont {Klimov}, \citenamefont {Knysh}, \citenamefont {Korotkov},
  \citenamefont {Kostritsa}, \citenamefont {Landhuis}, \citenamefont
  {Lindmark}, \citenamefont {Lucero}, \citenamefont {Lyakh}, \citenamefont
  {Mandr{\`a}}, \citenamefont {McClean}, \citenamefont {McEwen}, \citenamefont
  {Megrant}, \citenamefont {Mi}, \citenamefont {Michielsen}, \citenamefont
  {Mohseni}, \citenamefont {Mutus}, \citenamefont {Naaman}, \citenamefont
  {Neeley}, \citenamefont {Neill}, \citenamefont {Niu}, \citenamefont {Ostby},
  \citenamefont {Petukhov}, \citenamefont {Platt}, \citenamefont {Quintana},
  \citenamefont {Rieffel}, \citenamefont {Roushan}, \citenamefont {Rubin},
  \citenamefont {Sank}, \citenamefont {Satzinger}, \citenamefont {Smelyanskiy},
  \citenamefont {Sung}, \citenamefont {Trevithick}, \citenamefont
  {Vainsencher}, \citenamefont {Villalonga}, \citenamefont {White},
  \citenamefont {Yao}, \citenamefont {Yeh}, \citenamefont {Zalcman},
  \citenamefont {Neven},\ and\ \citenamefont {Martinis}}]{Arute2019}%
  \BibitemOpen
  \bibfield  {author} {\bibinfo {author} {\bibfnamefont {F.}~\bibnamefont
  {Arute}}, \bibinfo {author} {\bibfnamefont {K.}~\bibnamefont {Arya}},
  \bibinfo {author} {\bibfnamefont {R.}~\bibnamefont {Babbush}}, \bibinfo
  {author} {\bibfnamefont {D.}~\bibnamefont {Bacon}}, \bibinfo {author}
  {\bibfnamefont {J.~C.}\ \bibnamefont {Bardin}}, \bibinfo {author}
  {\bibfnamefont {R.}~\bibnamefont {Barends}}, \bibinfo {author} {\bibfnamefont
  {R.}~\bibnamefont {Biswas}}, \bibinfo {author} {\bibfnamefont
  {S.}~\bibnamefont {Boixo}}, \bibinfo {author} {\bibfnamefont {F.~G. S.~L.}\
  \bibnamefont {Brandao}}, \bibinfo {author} {\bibfnamefont {D.~A.}\
  \bibnamefont {Buell}}, \bibinfo {author} {\bibfnamefont {B.}~\bibnamefont
  {Burkett}}, \bibinfo {author} {\bibfnamefont {Y.}~\bibnamefont {Chen}},
  \bibinfo {author} {\bibfnamefont {Z.}~\bibnamefont {Chen}}, \bibinfo {author}
  {\bibfnamefont {B.}~\bibnamefont {Chiaro}}, \bibinfo {author} {\bibfnamefont
  {R.}~\bibnamefont {Collins}}, \bibinfo {author} {\bibfnamefont
  {W.}~\bibnamefont {Courtney}}, \bibinfo {author} {\bibfnamefont
  {A.}~\bibnamefont {Dunsworth}}, \bibinfo {author} {\bibfnamefont
  {E.}~\bibnamefont {Farhi}}, \bibinfo {author} {\bibfnamefont
  {B.}~\bibnamefont {Foxen}}, \bibinfo {author} {\bibfnamefont
  {A.}~\bibnamefont {Fowler}}, \bibinfo {author} {\bibfnamefont
  {C.}~\bibnamefont {Gidney}}, \bibinfo {author} {\bibfnamefont
  {M.}~\bibnamefont {Giustina}}, \bibinfo {author} {\bibfnamefont
  {R.}~\bibnamefont {Graff}}, \bibinfo {author} {\bibfnamefont
  {K.}~\bibnamefont {Guerin}}, \bibinfo {author} {\bibfnamefont
  {S.}~\bibnamefont {Habegger}}, \bibinfo {author} {\bibfnamefont {M.~P.}\
  \bibnamefont {Harrigan}}, \bibinfo {author} {\bibfnamefont {M.~J.}\
  \bibnamefont {Hartmann}}, \bibinfo {author} {\bibfnamefont {A.}~\bibnamefont
  {Ho}}, \bibinfo {author} {\bibfnamefont {M.}~\bibnamefont {Hoffmann}},
  \bibinfo {author} {\bibfnamefont {T.}~\bibnamefont {Huang}}, \bibinfo
  {author} {\bibfnamefont {T.~S.}\ \bibnamefont {Humble}}, \bibinfo {author}
  {\bibfnamefont {S.~V.}\ \bibnamefont {Isakov}}, \bibinfo {author}
  {\bibfnamefont {E.}~\bibnamefont {Jeffrey}}, \bibinfo {author} {\bibfnamefont
  {Z.}~\bibnamefont {Jiang}}, \bibinfo {author} {\bibfnamefont
  {D.}~\bibnamefont {Kafri}}, \bibinfo {author} {\bibfnamefont
  {K.}~\bibnamefont {Kechedzhi}}, \bibinfo {author} {\bibfnamefont
  {J.}~\bibnamefont {Kelly}}, \bibinfo {author} {\bibfnamefont {P.~V.}\
  \bibnamefont {Klimov}}, \bibinfo {author} {\bibfnamefont {S.}~\bibnamefont
  {Knysh}}, \bibinfo {author} {\bibfnamefont {A.}~\bibnamefont {Korotkov}},
  \bibinfo {author} {\bibfnamefont {F.}~\bibnamefont {Kostritsa}}, \bibinfo
  {author} {\bibfnamefont {D.}~\bibnamefont {Landhuis}}, \bibinfo {author}
  {\bibfnamefont {M.}~\bibnamefont {Lindmark}}, \bibinfo {author}
  {\bibfnamefont {E.}~\bibnamefont {Lucero}}, \bibinfo {author} {\bibfnamefont
  {D.}~\bibnamefont {Lyakh}}, \bibinfo {author} {\bibfnamefont
  {S.}~\bibnamefont {Mandr{\`a}}}, \bibinfo {author} {\bibfnamefont {J.~R.}\
  \bibnamefont {McClean}}, \bibinfo {author} {\bibfnamefont {M.}~\bibnamefont
  {McEwen}}, \bibinfo {author} {\bibfnamefont {A.}~\bibnamefont {Megrant}},
  \bibinfo {author} {\bibfnamefont {X.}~\bibnamefont {Mi}}, \bibinfo {author}
  {\bibfnamefont {K.}~\bibnamefont {Michielsen}}, \bibinfo {author}
  {\bibfnamefont {M.}~\bibnamefont {Mohseni}}, \bibinfo {author} {\bibfnamefont
  {J.}~\bibnamefont {Mutus}}, \bibinfo {author} {\bibfnamefont
  {O.}~\bibnamefont {Naaman}}, \bibinfo {author} {\bibfnamefont
  {M.}~\bibnamefont {Neeley}}, \bibinfo {author} {\bibfnamefont
  {C.}~\bibnamefont {Neill}}, \bibinfo {author} {\bibfnamefont {M.~Y.}\
  \bibnamefont {Niu}}, \bibinfo {author} {\bibfnamefont {E.}~\bibnamefont
  {Ostby}}, \bibinfo {author} {\bibfnamefont {A.}~\bibnamefont {Petukhov}},
  \bibinfo {author} {\bibfnamefont {J.~C.}\ \bibnamefont {Platt}}, \bibinfo
  {author} {\bibfnamefont {C.}~\bibnamefont {Quintana}}, \bibinfo {author}
  {\bibfnamefont {E.~G.}\ \bibnamefont {Rieffel}}, \bibinfo {author}
  {\bibfnamefont {P.}~\bibnamefont {Roushan}}, \bibinfo {author} {\bibfnamefont
  {N.~C.}\ \bibnamefont {Rubin}}, \bibinfo {author} {\bibfnamefont
  {D.}~\bibnamefont {Sank}}, \bibinfo {author} {\bibfnamefont {K.~J.}\
  \bibnamefont {Satzinger}}, \bibinfo {author} {\bibfnamefont {V.}~\bibnamefont
  {Smelyanskiy}}, \bibinfo {author} {\bibfnamefont {K.~J.}\ \bibnamefont
  {Sung}}, \bibinfo {author} {\bibfnamefont {M.~D.}\ \bibnamefont
  {Trevithick}}, \bibinfo {author} {\bibfnamefont {A.}~\bibnamefont
  {Vainsencher}}, \bibinfo {author} {\bibfnamefont {B.}~\bibnamefont
  {Villalonga}}, \bibinfo {author} {\bibfnamefont {T.}~\bibnamefont {White}},
  \bibinfo {author} {\bibfnamefont {Z.~J.}\ \bibnamefont {Yao}}, \bibinfo
  {author} {\bibfnamefont {P.}~\bibnamefont {Yeh}}, \bibinfo {author}
  {\bibfnamefont {A.}~\bibnamefont {Zalcman}}, \bibinfo {author} {\bibfnamefont
  {H.}~\bibnamefont {Neven}}, \ and\ \bibinfo {author} {\bibfnamefont {J.~M.}\
  \bibnamefont {Martinis}},\ }\href {\doibase 10.1038/s41586-019-1666-5}
  {\bibfield  {journal} {\bibinfo  {journal} {Nature}\ }\textbf {\bibinfo
  {volume} {574}},\ \bibinfo {pages} {505} (\bibinfo {year}
  {2019})}\BibitemShut {NoStop}%
\bibitem [{\citenamefont {Verstraete}\ and\ \citenamefont
  {Cirac}(2004)}]{Verstraete2004}%
  \BibitemOpen
  \bibfield  {author} {\bibinfo {author} {\bibfnamefont {F.}~\bibnamefont
  {Verstraete}}\ and\ \bibinfo {author} {\bibfnamefont {J.~I.}\ \bibnamefont
  {Cirac}},\ }\href {http://arxiv.org/abs/cond-mat/0407066} {\bibfield
  {journal} {\bibinfo  {journal} {Annalen der Physik}\ }\textbf {\bibinfo
  {volume} {4}},\ \bibinfo {pages} {1} (\bibinfo {year} {2004})},\ \Eprint
  {http://arxiv.org/abs/cond-mat/0407066} {arXiv:cond-mat/0407066 [cond-mat]}
  \BibitemShut {NoStop}%
\bibitem [{\citenamefont {Saadatmand}\ and\ \citenamefont
  {McCulloch}(2016)}]{Saadatmand2016}%
  \BibitemOpen
  \bibfield  {author} {\bibinfo {author} {\bibfnamefont {S.~N.}\ \bibnamefont
  {Saadatmand}}\ and\ \bibinfo {author} {\bibfnamefont {I.~P.}\ \bibnamefont
  {McCulloch}},\ }\href {\doibase 10.1103/PhysRevB.94.121111} {\bibfield
  {journal} {\bibinfo  {journal} {Physical Review B}\ }\textbf {\bibinfo
  {volume} {94}},\ \bibinfo {pages} {121111} (\bibinfo {year}
  {2016})}\BibitemShut {NoStop}%
\bibitem [{\citenamefont {Haegeman}\ \emph {et~al.}(2016)\citenamefont
  {Haegeman}, \citenamefont {Lubich}, \citenamefont {Oseledets}, \citenamefont
  {Vandereycken},\ and\ \citenamefont {Verstraete}}]{Haegeman2016}%
  \BibitemOpen
  \bibfield  {author} {\bibinfo {author} {\bibfnamefont {J.}~\bibnamefont
  {Haegeman}}, \bibinfo {author} {\bibfnamefont {C.}~\bibnamefont {Lubich}},
  \bibinfo {author} {\bibfnamefont {I.}~\bibnamefont {Oseledets}}, \bibinfo
  {author} {\bibfnamefont {B.}~\bibnamefont {Vandereycken}}, \ and\ \bibinfo
  {author} {\bibfnamefont {F.}~\bibnamefont {Verstraete}},\ }\href {\doibase
  10.1103/PhysRevB.94.165116} {\bibfield  {journal} {\bibinfo  {journal}
  {Physical Review B}\ }\textbf {\bibinfo {volume} {94}},\ \bibinfo {pages}
  {165116} (\bibinfo {year} {2016})}\BibitemShut {NoStop}%
\bibitem [{\citenamefont {Zauner-Stauber}\ \emph {et~al.}(2018)\citenamefont
  {Zauner-Stauber}, \citenamefont {Vanderstraeten}, \citenamefont {Fishman},
  \citenamefont {Verstraete},\ and\ \citenamefont {Haegeman}}]{Zauner2018}%
  \BibitemOpen
  \bibfield  {author} {\bibinfo {author} {\bibfnamefont {V.}~\bibnamefont
  {Zauner-Stauber}}, \bibinfo {author} {\bibfnamefont {L.}~\bibnamefont
  {Vanderstraeten}}, \bibinfo {author} {\bibfnamefont {M.~T.}\ \bibnamefont
  {Fishman}}, \bibinfo {author} {\bibfnamefont {F.}~\bibnamefont {Verstraete}},
  \ and\ \bibinfo {author} {\bibfnamefont {J.}~\bibnamefont {Haegeman}},\
  }\href {\doibase 10.1103/PhysRevB.97.045145} {\bibfield  {journal} {\bibinfo
  {journal} {Phys. Rev. B}\ }\textbf {\bibinfo {volume} {97}},\ \bibinfo
  {pages} {045145} (\bibinfo {year} {2018})}\BibitemShut {NoStop}%
\bibitem [{\citenamefont {Schollw{\"{o}}ck}(2011)}]{Schollwock2011}%
  \BibitemOpen
  \bibfield  {author} {\bibinfo {author} {\bibfnamefont {U.}~\bibnamefont
  {Schollw{\"{o}}ck}},\ }\href {\doibase 10.1016/j.aop.2010.09.012} {\bibfield
  {journal} {\bibinfo  {journal} {Annals of Physics}\ }\textbf {\bibinfo
  {volume} {326}},\ \bibinfo {pages} {96} (\bibinfo {year} {2011})},\ \Eprint
  {http://arxiv.org/abs/1008.3477} {arXiv:1008.3477} \BibitemShut {NoStop}%
\bibitem [{\citenamefont {McCulloch}(2008{\natexlab{b}})}]{McCulloch2008}%
  \BibitemOpen
  \bibfield  {author} {\bibinfo {author} {\bibfnamefont {I.}~\bibnamefont
  {McCulloch}},\ }\href {http://arxiv.org/abs/0804.2509} {\bibfield  {journal}
  {\bibinfo  {journal} {Arxiv preprint arXiv:0804.2509}\ ,\ \bibinfo {pages}
  {12}} (\bibinfo {year} {2008}{\natexlab{b}})},\ \Eprint
  {http://arxiv.org/abs/0804.2509} {arXiv:0804.2509} \BibitemShut {NoStop}%
\bibitem [{\citenamefont {Vidal}(2007)}]{Vidal2007}%
  \BibitemOpen
  \bibfield  {author} {\bibinfo {author} {\bibfnamefont {G.}~\bibnamefont
  {Vidal}},\ }\href {\doibase 10.1103/PhysRevLett.98.070201} {\bibfield
  {journal} {\bibinfo  {journal} {Physical Review Letters}\ }\textbf {\bibinfo
  {volume} {98}},\ \bibinfo {pages} {070201} (\bibinfo {year}
  {2007})}\BibitemShut {NoStop}%
\bibitem [{\citenamefont {Paeckel}\ \emph {et~al.}(2019)\citenamefont
  {Paeckel}, \citenamefont {K{\"{o}}hler}, \citenamefont {Swoboda},
  \citenamefont {Manmana}, \citenamefont {Schollw{\"{o}}ck},\ and\
  \citenamefont {Hubig}}]{Paeckel2019}%
  \BibitemOpen
  \bibfield  {author} {\bibinfo {author} {\bibfnamefont {S.}~\bibnamefont
  {Paeckel}}, \bibinfo {author} {\bibfnamefont {T.}~\bibnamefont
  {K{\"{o}}hler}}, \bibinfo {author} {\bibfnamefont {A.}~\bibnamefont
  {Swoboda}}, \bibinfo {author} {\bibfnamefont {S.~R.}\ \bibnamefont
  {Manmana}}, \bibinfo {author} {\bibfnamefont {U.}~\bibnamefont
  {Schollw{\"{o}}ck}}, \ and\ \bibinfo {author} {\bibfnamefont
  {C.}~\bibnamefont {Hubig}},\ }\href {http://arxiv.org/abs/1901.05824} {\
  (\bibinfo {year} {2019})},\ \Eprint {http://arxiv.org/abs/1901.05824}
  {arXiv:1901.05824} \BibitemShut {NoStop}%
\bibitem [{\citenamefont {Chan}\ \emph {et~al.}(2016)\citenamefont {Chan},
  \citenamefont {Keselman}, \citenamefont {Nakatani}, \citenamefont {Li},\ and\
  \citenamefont {White}}]{Chan2016}%
  \BibitemOpen
  \bibfield  {author} {\bibinfo {author} {\bibfnamefont {G.~K.-L.}\
  \bibnamefont {Chan}}, \bibinfo {author} {\bibfnamefont {A.}~\bibnamefont
  {Keselman}}, \bibinfo {author} {\bibfnamefont {N.}~\bibnamefont {Nakatani}},
  \bibinfo {author} {\bibfnamefont {Z.}~\bibnamefont {Li}}, \ and\ \bibinfo
  {author} {\bibfnamefont {S.~R.}\ \bibnamefont {White}},\ }\href {\doibase
  10.1063/1.4955108} {\bibfield  {journal} {\bibinfo  {journal} {The Journal of
  Chemical Physics}\ }\textbf {\bibinfo {volume} {145}},\ \bibinfo {pages}
  {014102} (\bibinfo {year} {2016})},\ \Eprint
  {http://arxiv.org/abs/1605.02611} {arXiv:1605.02611} \BibitemShut {NoStop}%
\bibitem [{\citenamefont {Pirvu}\ \emph {et~al.}(2010)\citenamefont {Pirvu},
  \citenamefont {Murg}, \citenamefont {Cirac},\ and\ \citenamefont
  {Verstraete}}]{Pirvu2010}%
  \BibitemOpen
  \bibfield  {author} {\bibinfo {author} {\bibfnamefont {B.}~\bibnamefont
  {Pirvu}}, \bibinfo {author} {\bibfnamefont {V.}~\bibnamefont {Murg}},
  \bibinfo {author} {\bibfnamefont {J.~I.}\ \bibnamefont {Cirac}}, \ and\
  \bibinfo {author} {\bibfnamefont {F.}~\bibnamefont {Verstraete}},\ }\href
  {\doibase 10.1088/1367-2630/12/2/025012} {\bibfield  {journal} {\bibinfo
  {journal} {New Journal of Physics}\ }\textbf {\bibinfo {volume} {12}},\
  \bibinfo {pages} {025012} (\bibinfo {year} {2010})},\ \Eprint
  {http://arxiv.org/abs/0804.3976} {arXiv:0804.3976} \BibitemShut {NoStop}%
\bibitem [{\citenamefont {Suzuki}(1985)}]{Suzuki1985}%
  \BibitemOpen
  \bibfield  {author} {\bibinfo {author} {\bibfnamefont {M.}~\bibnamefont
  {Suzuki}},\ }\href {\doibase 10.1016/0375-9601(85)90168-9} {\bibfield
  {journal} {\bibinfo  {journal} {Physics Letters A}\ }\textbf {\bibinfo
  {volume} {113}},\ \bibinfo {pages} {299} (\bibinfo {year}
  {1985})}\BibitemShut {NoStop}%
\bibitem [{\citenamefont {Or{\'{u}}s}\ and\ \citenamefont
  {Vidal}(2008)}]{Orus2007}%
  \BibitemOpen
  \bibfield  {author} {\bibinfo {author} {\bibfnamefont {R.}~\bibnamefont
  {Or{\'{u}}s}}\ and\ \bibinfo {author} {\bibfnamefont {G.}~\bibnamefont
  {Vidal}},\ }\href {\doibase 10.1103/PhysRevB.78.155117} {\bibfield  {journal}
  {\bibinfo  {journal} {Physical Review B}\ }\textbf {\bibinfo {volume} {78}},\
  \bibinfo {pages} {155117} (\bibinfo {year} {2008})},\ \Eprint
  {http://arxiv.org/abs/0711.3960} {arXiv:0711.3960} \BibitemShut {NoStop}%
\bibitem [{\citenamefont {Krylov}(1931)}]{krylov1931resolution}%
  \BibitemOpen
  \bibfield  {author} {\bibinfo {author} {\bibfnamefont {A.~N.}\ \bibnamefont
  {Krylov}},\ }\href@noop {} {\bibfield  {journal} {\bibinfo  {journal}
  {Izvestiya Rossiiskoi Akademii Nauk. Seriya Matematicheskaya}\ ,\ \bibinfo
  {pages} {491}} (\bibinfo {year} {1931})}\BibitemShut {NoStop}%
\bibitem [{\citenamefont {Moler}\ and\ \citenamefont {{Van
  Loan}}(2003)}]{Moler2003}%
  \BibitemOpen
  \bibfield  {author} {\bibinfo {author} {\bibfnamefont {C.}~\bibnamefont
  {Moler}}\ and\ \bibinfo {author} {\bibfnamefont {C.}~\bibnamefont {{Van
  Loan}}},\ }\href {\doibase 10.1137/S00361445024180} {\bibfield  {journal}
  {\bibinfo  {journal} {SIAM Review}\ }\textbf {\bibinfo {volume} {45}},\
  \bibinfo {pages} {3} (\bibinfo {year} {2003})}\BibitemShut {NoStop}%
\bibitem [{\citenamefont {Lanczos}(1950)}]{Lanczos1950}%
  \BibitemOpen
  \bibfield  {author} {\bibinfo {author} {\bibfnamefont {C.}~\bibnamefont
  {Lanczos}},\ }\href {\doibase 10.6028/jres.045.026} {\bibfield  {journal}
  {\bibinfo  {journal} {Journal of Research of the National Bureau of
  Standards}\ }\textbf {\bibinfo {volume} {45}},\ \bibinfo {pages} {255}
  (\bibinfo {year} {1950})}\BibitemShut {NoStop}%
\bibitem [{\citenamefont {Stewart}(2002)}]{Stewart2002}%
  \BibitemOpen
  \bibfield  {author} {\bibinfo {author} {\bibfnamefont {G.~W.}\ \bibnamefont
  {Stewart}},\ }\href {\doibase 10.1137/S1064827501388984} {\bibfield
  {journal} {\bibinfo  {journal} {SIAM Journal on Scientific Computing}\
  }\textbf {\bibinfo {volume} {24}},\ \bibinfo {pages} {201} (\bibinfo {year}
  {2002})}\BibitemShut {NoStop}%
\bibitem [{\citenamefont {Orecchia}\ \emph {et~al.}(2012)\citenamefont
  {Orecchia}, \citenamefont {Sachdeva},\ and\ \citenamefont
  {Vishnoi}}]{Orecchia2012}%
  \BibitemOpen
  \bibfield  {author} {\bibinfo {author} {\bibfnamefont {L.}~\bibnamefont
  {Orecchia}}, \bibinfo {author} {\bibfnamefont {S.}~\bibnamefont {Sachdeva}},
  \ and\ \bibinfo {author} {\bibfnamefont {N.~K.}\ \bibnamefont {Vishnoi}},\
  }in\ \href {\doibase 10.1145/2213977.2214080} {\emph {\bibinfo {booktitle}
  {Proceedings of the 44th symposium on Theory of Computing - STOC '12}}}\
  (\bibinfo  {publisher} {ACM Press},\ \bibinfo {address} {New York, New York,
  USA},\ \bibinfo {year} {2012})\ p.\ \bibinfo {pages} {1141}\BibitemShut
  {NoStop}%
\bibitem [{\citenamefont {Saad}(1992)}]{Saad}%
  \BibitemOpen
  \bibfield  {author} {\bibinfo {author} {\bibfnamefont {Y.}~\bibnamefont
  {Saad}},\ }\href {\doibase 10.1137/0729014} {\bibfield  {journal} {\bibinfo
  {journal} {SIAM Journal on Numerical Analysis}\ }\textbf {\bibinfo {volume}
  {29}},\ \bibinfo {pages} {209} (\bibinfo {year} {1992})}\BibitemShut
  {NoStop}%
\bibitem [{\citenamefont {Koskela}\ and\ \citenamefont
  {Ostermann}(2013)}]{Koskela2013}%
  \BibitemOpen
  \bibfield  {author} {\bibinfo {author} {\bibfnamefont {A.}~\bibnamefont
  {Koskela}}\ and\ \bibinfo {author} {\bibfnamefont {A.}~\bibnamefont
  {Ostermann}},\ }\href {\doibase 10.1016/j.camwa.2012.06.004} {\bibfield
  {journal} {\bibinfo  {journal} {Computers {\&} Mathematics with
  Applications}\ }\textbf {\bibinfo {volume} {65}},\ \bibinfo {pages} {487}
  (\bibinfo {year} {2013})}\BibitemShut {NoStop}%
\bibitem [{\citenamefont {Simon}(1984)}]{Simon1984}%
  \BibitemOpen
  \bibfield  {author} {\bibinfo {author} {\bibfnamefont {H.~D.}\ \bibnamefont
  {Simon}},\ }\href {\doibase 10.1016/0024-3795(84)90025-9} {\bibfield
  {journal} {\bibinfo  {journal} {Linear Algebra and its Applications}\
  }\textbf {\bibinfo {volume} {61}},\ \bibinfo {pages} {101} (\bibinfo {year}
  {1984})}\BibitemShut {NoStop}%
\bibitem [{\citenamefont {Haegeman}\ \emph {et~al.}(2013)\citenamefont
  {Haegeman}, \citenamefont {Osborne},\ and\ \citenamefont
  {Verstraete}}]{Haegeman2013}%
  \BibitemOpen
  \bibfield  {author} {\bibinfo {author} {\bibfnamefont {J.}~\bibnamefont
  {Haegeman}}, \bibinfo {author} {\bibfnamefont {T.~J.}\ \bibnamefont
  {Osborne}}, \ and\ \bibinfo {author} {\bibfnamefont {F.}~\bibnamefont
  {Verstraete}},\ }\href {\doibase 10.1103/PhysRevB.88.075133} {\bibfield
  {journal} {\bibinfo  {journal} {Phys. Rev. B}\ }\textbf {\bibinfo {volume}
  {88}},\ \bibinfo {pages} {075133} (\bibinfo {year} {2013})}\BibitemShut
  {NoStop}%
\bibitem [{mpt()}]{mptoolkit}%
  \BibitemOpen
  \href@noop {} {}\bibinfo {howpublished} {I.~P.~McCulloch, \textit{Matrix
  Product Toolkit}. URL:
  https://people.smp.uq.edu.au/IanMcCulloch/mptoolkit/}\BibitemShut {NoStop}%
\bibitem [{Has()}]{Hashizume_thesis}%
  \BibitemOpen
  \href@noop {} {}\bibinfo {howpublished} {Tomohiro Hashizume, Honours Thesis,
  University of Queensland, 2017}\BibitemShut {NoStop}%
\bibitem [{\citenamefont {Andraschko}\ and\ \citenamefont
  {Sirker}(2014)}]{Andraschko2014}%
  \BibitemOpen
  \bibfield  {author} {\bibinfo {author} {\bibfnamefont {F.}~\bibnamefont
  {Andraschko}}\ and\ \bibinfo {author} {\bibfnamefont {J.}~\bibnamefont
  {Sirker}},\ }\href {\doibase 10.1103/PhysRevB.89.125120} {\bibfield
  {journal} {\bibinfo  {journal} {Phys. Rev. B}\ }\textbf {\bibinfo {volume}
  {89}},\ \bibinfo {pages} {125120} (\bibinfo {year} {2014})}\BibitemShut
  {NoStop}%
\bibitem [{\citenamefont {Vajna}\ and\ \citenamefont
  {D\'ora}(2014)}]{Vajna2014}%
  \BibitemOpen
  \bibfield  {author} {\bibinfo {author} {\bibfnamefont {S.}~\bibnamefont
  {Vajna}}\ and\ \bibinfo {author} {\bibfnamefont {B.}~\bibnamefont {D\'ora}},\
  }\href {\doibase 10.1103/PhysRevB.89.161105} {\bibfield  {journal} {\bibinfo
  {journal} {Phys. Rev. B}\ }\textbf {\bibinfo {volume} {89}},\ \bibinfo
  {pages} {161105} (\bibinfo {year} {2014})}\BibitemShut {NoStop}%
\bibitem [{\citenamefont {Halimeh}\ and\ \citenamefont
  {Zauner-Stauber}(2017)}]{Halimeh2017}%
  \BibitemOpen
  \bibfield  {author} {\bibinfo {author} {\bibfnamefont {J.~C.}\ \bibnamefont
  {Halimeh}}\ and\ \bibinfo {author} {\bibfnamefont {V.}~\bibnamefont
  {Zauner-Stauber}},\ }\href {\doibase 10.1103/PhysRevB.96.134427} {\bibfield
  {journal} {\bibinfo  {journal} {Physical Review B}\ }\textbf {\bibinfo
  {volume} {96}},\ \bibinfo {pages} {134427} (\bibinfo {year} {2017})},\
  \Eprint {http://arxiv.org/abs/1610.02019} {arXiv:1610.02019} \BibitemShut
  {NoStop}%
\bibitem [{\citenamefont {Zauner-Stauber}\ and\ \citenamefont
  {Halimeh}(2017)}]{Zauner2017}%
  \BibitemOpen
  \bibfield  {author} {\bibinfo {author} {\bibfnamefont {V.}~\bibnamefont
  {Zauner-Stauber}}\ and\ \bibinfo {author} {\bibfnamefont {J.~C.}\
  \bibnamefont {Halimeh}},\ }\href {\doibase 10.1103/PhysRevE.96.062118}
  {\bibfield  {journal} {\bibinfo  {journal} {Phys. Rev. E}\ }\textbf {\bibinfo
  {volume} {96}},\ \bibinfo {pages} {062118} (\bibinfo {year}
  {2017})}\BibitemShut {NoStop}%
\bibitem [{\citenamefont {Homrighausen}\ \emph {et~al.}(2017)\citenamefont
  {Homrighausen}, \citenamefont {Abeling}, \citenamefont {Zauner-Stauber},\
  and\ \citenamefont {Halimeh}}]{Homrighausen2017}%
  \BibitemOpen
  \bibfield  {author} {\bibinfo {author} {\bibfnamefont {I.}~\bibnamefont
  {Homrighausen}}, \bibinfo {author} {\bibfnamefont {N.~O.}\ \bibnamefont
  {Abeling}}, \bibinfo {author} {\bibfnamefont {V.}~\bibnamefont
  {Zauner-Stauber}}, \ and\ \bibinfo {author} {\bibfnamefont {J.~C.}\
  \bibnamefont {Halimeh}},\ }\href {\doibase 10.1103/PhysRevB.96.104436}
  {\bibfield  {journal} {\bibinfo  {journal} {Phys. Rev. B}\ }\textbf {\bibinfo
  {volume} {96}},\ \bibinfo {pages} {104436} (\bibinfo {year}
  {2017})}\BibitemShut {NoStop}%
\bibitem [{\citenamefont {Lang}\ \emph
  {et~al.}(2018{\natexlab{a}})\citenamefont {Lang}, \citenamefont {Frank},\
  and\ \citenamefont {Halimeh}}]{Lang2017}%
  \BibitemOpen
  \bibfield  {author} {\bibinfo {author} {\bibfnamefont {J.}~\bibnamefont
  {Lang}}, \bibinfo {author} {\bibfnamefont {B.}~\bibnamefont {Frank}}, \ and\
  \bibinfo {author} {\bibfnamefont {J.~C.}\ \bibnamefont {Halimeh}},\ }\href
  {\doibase 10.1103/PhysRevB.97.174401} {\bibfield  {journal} {\bibinfo
  {journal} {Phys. Rev. B}\ }\textbf {\bibinfo {volume} {97}},\ \bibinfo
  {pages} {174401} (\bibinfo {year} {2018}{\natexlab{a}})}\BibitemShut
  {NoStop}%
\bibitem [{\citenamefont {Lang}\ \emph
  {et~al.}(2018{\natexlab{b}})\citenamefont {Lang}, \citenamefont {Frank},\
  and\ \citenamefont {Halimeh}}]{Lang2018}%
  \BibitemOpen
  \bibfield  {author} {\bibinfo {author} {\bibfnamefont {J.}~\bibnamefont
  {Lang}}, \bibinfo {author} {\bibfnamefont {B.}~\bibnamefont {Frank}}, \ and\
  \bibinfo {author} {\bibfnamefont {J.~C.}\ \bibnamefont {Halimeh}},\ }\href
  {\doibase 10.1103/PhysRevLett.121.130603} {\bibfield  {journal} {\bibinfo
  {journal} {Phys. Rev. Lett.}\ }\textbf {\bibinfo {volume} {121}},\ \bibinfo
  {pages} {130603} (\bibinfo {year} {2018}{\natexlab{b}})}\BibitemShut
  {NoStop}%
\bibitem [{\citenamefont {{Hashizume}}\ \emph {et~al.}(2018)\citenamefont
  {{Hashizume}}, \citenamefont {{McCulloch}},\ and\ \citenamefont
  {{Halimeh}}}]{Hashizume2018}%
  \BibitemOpen
  \bibfield  {author} {\bibinfo {author} {\bibfnamefont {T.}~\bibnamefont
  {{Hashizume}}}, \bibinfo {author} {\bibfnamefont {I.~P.}\ \bibnamefont
  {{McCulloch}}}, \ and\ \bibinfo {author} {\bibfnamefont {J.~C.}\ \bibnamefont
  {{Halimeh}}},\ }\href {https://arxiv.org/abs/1811.09275} {\bibfield
  {journal} {\bibinfo  {journal} {ArXiv e-prints}\ } (\bibinfo {year}
  {2018})},\ \Eprint {http://arxiv.org/abs/1811.09275} {arXiv:1811.09275
  [cond-mat.str-el]} \BibitemShut {NoStop}%
\bibitem [{\citenamefont {{Halimeh}}\ \emph {et~al.}(2018)\citenamefont
  {{Halimeh}}, \citenamefont {{Van Damme}}, \citenamefont {{Zauner-Stauber}},\
  and\ \citenamefont {{Vanderstraeten}}}]{Halimeh2018a}%
  \BibitemOpen
  \bibfield  {author} {\bibinfo {author} {\bibfnamefont {J.~C.}\ \bibnamefont
  {{Halimeh}}}, \bibinfo {author} {\bibfnamefont {M.}~\bibnamefont {{Van
  Damme}}}, \bibinfo {author} {\bibfnamefont {V.}~\bibnamefont
  {{Zauner-Stauber}}}, \ and\ \bibinfo {author} {\bibfnamefont
  {L.}~\bibnamefont {{Vanderstraeten}}},\ }\href
  {https://arxiv.org/abs/1810.07187} {\bibfield  {journal} {\bibinfo  {journal}
  {ArXiv e-prints}\ } (\bibinfo {year} {2018})},\ \Eprint
  {http://arxiv.org/abs/1810.07187} {arXiv:1810.07187 [cond-mat.str-el]}
  \BibitemShut {NoStop}%
\bibitem [{\citenamefont {Defenu}\ \emph {et~al.}(2019)\citenamefont {Defenu},
  \citenamefont {Enss},\ and\ \citenamefont {Halimeh}}]{Defenu2019}%
  \BibitemOpen
  \bibfield  {author} {\bibinfo {author} {\bibfnamefont {N.}~\bibnamefont
  {Defenu}}, \bibinfo {author} {\bibfnamefont {T.}~\bibnamefont {Enss}}, \ and\
  \bibinfo {author} {\bibfnamefont {J.~C.}\ \bibnamefont {Halimeh}},\ }\href
  {\doibase 10.1103/PhysRevB.100.014434} {\bibfield  {journal} {\bibinfo
  {journal} {Phys. Rev. B}\ }\textbf {\bibinfo {volume} {100}},\ \bibinfo
  {pages} {014434} (\bibinfo {year} {2019})}\BibitemShut {NoStop}%
\bibitem [{\citenamefont {Schmitt}\ and\ \citenamefont
  {Kehrein}(2015)}]{Schmitt2015}%
  \BibitemOpen
  \bibfield  {author} {\bibinfo {author} {\bibfnamefont {M.}~\bibnamefont
  {Schmitt}}\ and\ \bibinfo {author} {\bibfnamefont {S.}~\bibnamefont
  {Kehrein}},\ }\href {\doibase 10.1103/PhysRevB.92.075114} {\bibfield
  {journal} {\bibinfo  {journal} {Phys. Rev. B}\ }\textbf {\bibinfo {volume}
  {92}},\ \bibinfo {pages} {075114} (\bibinfo {year} {2015})}\BibitemShut
  {NoStop}%
\bibitem [{\citenamefont {Bhattacharya}\ and\ \citenamefont
  {Dutta}(2017)}]{Bhattacharya2017}%
  \BibitemOpen
  \bibfield  {author} {\bibinfo {author} {\bibfnamefont {U.}~\bibnamefont
  {Bhattacharya}}\ and\ \bibinfo {author} {\bibfnamefont {A.}~\bibnamefont
  {Dutta}},\ }\href {\doibase 10.1103/PhysRevB.95.184307} {\bibfield  {journal}
  {\bibinfo  {journal} {Phys. Rev. B}\ }\textbf {\bibinfo {volume} {95}},\
  \bibinfo {pages} {184307} (\bibinfo {year} {2017})}\BibitemShut {NoStop}%
\bibitem [{\citenamefont {Karrasch}\ and\ \citenamefont
  {Schuricht}(2013)}]{Karrasch2013}%
  \BibitemOpen
  \bibfield  {author} {\bibinfo {author} {\bibfnamefont {C.}~\bibnamefont
  {Karrasch}}\ and\ \bibinfo {author} {\bibfnamefont {D.}~\bibnamefont
  {Schuricht}},\ }\href {\doibase 10.1103/PhysRevB.87.195104} {\bibfield
  {journal} {\bibinfo  {journal} {Physical Review B}\ }\textbf {\bibinfo
  {volume} {87}},\ \bibinfo {pages} {195104} (\bibinfo {year} {2013})},\
  \Eprint {http://arxiv.org/abs/arXiv:1302.3893v2} {arXiv:arXiv:1302.3893v2}
  \BibitemShut {NoStop}%
\bibitem [{\citenamefont {Saadatmand}\ \emph {et~al.}(2018)\citenamefont
  {Saadatmand}, \citenamefont {Bartlett},\ and\ \citenamefont
  {McCulloch}}]{Saadatmand2018}%
  \BibitemOpen
  \bibfield  {author} {\bibinfo {author} {\bibfnamefont {S.~N.}\ \bibnamefont
  {Saadatmand}}, \bibinfo {author} {\bibfnamefont {S.~D.}\ \bibnamefont
  {Bartlett}}, \ and\ \bibinfo {author} {\bibfnamefont {I.~P.}\ \bibnamefont
  {McCulloch}},\ }\href {\doibase 10.1103/PhysRevB.97.155116} {\bibfield
  {journal} {\bibinfo  {journal} {Phys. Rev. B}\ }\textbf {\bibinfo {volume}
  {97}},\ \bibinfo {pages} {155116} (\bibinfo {year} {2018})}\BibitemShut
  {NoStop}%
\end{thebibliography}%
\end{document}